\documentclass[a4paper,11pt]{article}
\usepackage{jcappub}
\usepackage{aas_macros}
\usepackage{lineno}
\usepackage{orcidlink}
\usepackage{bm}
\usepackage{mathrsfs}

\newcommand{\lya}{Ly$\alpha$}
\newcommand{\poned}{$P_{\mathrm{1D}}$}
\newcommand{\skm}{s~km$^{-1}$}
\newcommand{\kms}{km~s$^{-1}$}
\newcommand{\impc}{Mpc$^{-1}$}
\newcommand{\Tr}{\mathrm{Tr}}
\newcommand{\qsonic}{\texttt{qsonic}}

\title{\boldmath DESI DR1 Ly$\alpha$ 1D power spectrum: The optimal estimator measurement}

\author[1,2,3]{{Naim~G\" oksel Kara\c{c}ayl{\i}}\orcidlink{0000-0001-7336-8912},}
\author[1,2]{{Paul Martini}\orcidlink{0000-0002-4279-4182},}
\author[4]{{J.~Aguilar},}
\author[5]{{S.~Ahlen}\orcidlink{0000-0001-6098-7247},}
\author[6]{{E.~Armengaud}\orcidlink{0000-0001-7600-5148},}
\author[4]{{S.~Bailey}\orcidlink{0000-0003-4162-6619},}
\author[4]{{A.~Bault}\orcidlink{0000-0002-9964-1005},}
\author[7,8]{{D.~Bianchi}\orcidlink{0000-0001-9712-0006},}
\author[4]{{A.~Brodzeller}\orcidlink{0000-0002-8934-0954},}
\author[9]{{D.~Brooks},}
\author[10]{{J.~Chaves-Montero}\orcidlink{0000-0002-9553-4261},}
\author[4]{{T.~Claybaugh},}
\author[4,11]{{A.~Cuceu}\orcidlink{0000-0002-2169-0595},}
\author[12]{{A.~de~la~Macorra}\orcidlink{0000-0002-1769-1640},}
\author[13]{{A.~Dey}\orcidlink{0000-0002-4928-4003},}
\author[14,15]{{B.~Dey}\orcidlink{0000-0002-5665-7912},}
\author[9]{{P.~Doel},}
\author[4,16]{{S.~Ferraro}\orcidlink{0000-0003-4992-7854},}
\author[10]{{A.~Font-Ribera}\orcidlink{0000-0002-3033-7312},}
\author[17,18]{{J.~E.~Forero-Romero}\orcidlink{0000-0002-2890-3725},}
\author[19,20,21]{{E.~Gaztañaga},}
\author[4]{{S.~Gontcho A Gontcho}\orcidlink{0000-0003-3142-233X},}
\author[22]{{G.~Gutierrez},}
\author[4]{{J.~Guy}\orcidlink{0000-0001-9822-6793},}
\author[23,23]{{C.~Hahn}\orcidlink{0000-0003-1197-0902},}
\author[24,6]{{H.~K.~Herrera-Alcantar}\orcidlink{0000-0002-9136-9609},}
\author[1,3]{{K.~Honscheid}\orcidlink{0000-0002-6550-2023},}
\author[25]{{M.~Ishak}\orcidlink{0000-0002-6024-466X},}
\author[26]{{R.~Kehoe},}
\author[27]{{D.~Kirkby}\orcidlink{0000-0002-8828-5463},}
\author[4]{{A.~Kremin}\orcidlink{0000-0001-6356-7424},}
\author[4]{{M.~Landriau}\orcidlink{0000-0003-1838-8528},}
\author[6]{{J.~M.~Le~Goff},}
\author[28]{{L.~Le~Guillou}\orcidlink{0000-0001-7178-8868},}
\author[4]{{M.~E.~Levi}\orcidlink{0000-0003-1887-1018},}
\author[29,10]{{M.~Manera}\orcidlink{0000-0003-4962-8934},}
\author[13]{{A.~Meisner}\orcidlink{0000-0002-1125-7384},}
\author[30,10]{{R.~Miquel},}
\author[31]{{P.~Montero-Camacho}\orcidlink{0000-0002-6998-6678},}
\author[20]{{S.~Nadathur}\orcidlink{0000-0001-9070-3102},}
\author[32,33]{{G.~Niz}\orcidlink{0000-0002-1544-8946},}
\author[6,4]{{N.~Palanque-Delabrouille}\orcidlink{0000-0003-3188-784X},}
\author[34]{{Z.~Pan}\orcidlink{0000-0003-0230-6436},}
\author[35,36,37]{{W.~J.~Percival}\orcidlink{0000-0002-0644-5727},}
\author[38]{{Matthew~M.~Pieri}\orcidlink{0000-0003-0247-8991},}
\author[39]{{F.~Prada}\orcidlink{0000-0001-7145-8674},}
\author[40]{{I.~P\'erez-R\`afols}\orcidlink{0000-0001-6979-0125},}
\author[41]{{C.~Ravoux}\orcidlink{0000-0002-3500-6635},}
\author[42]{{G.~Rossi},}
\author[43]{{E.~Sanchez}\orcidlink{0000-0002-9646-8198},}
\author[44]{{C.~Saulder}\orcidlink{0000-0002-0408-5633},}
\author[4]{{D.~Schlegel},}
\author[45]{{M.~Schubnell},}
\author[46]{{H.~Seo}\orcidlink{0000-0002-6588-3508},}
\author[21,47]{{M.~Siudek}\orcidlink{0000-0002-2949-2155},}
\author[13]{{D.~Sprayberry},}
\author[6]{{T.~Tan}\orcidlink{0000-0001-8289-1481},}
\author[48]{{Ji-Jia Tang}\orcidlink{0000-0002-1860-0886},}
\author[45]{{G.~Tarl\'{e}}\orcidlink{0000-0003-1704-0781},}
\author[49,50]{{M.~Walther}\orcidlink{0000-0002-1748-3745},}
\author[13]{{B.~A.~Weaver},}
\author[51]{{J.~Yu}\orcidlink{0009-0001-7217-8006},}
\author[4]{{R.~Zhou}\orcidlink{0000-0001-5381-4372},}
\author[52]{{H.~Zou}\orcidlink{0000-0002-6684-3997},}

\affiliation[1]{Center for Cosmology and AstroParticle Physics, The Ohio State University, 191 West Woodruff Avenue, Columbus, OH 43210, USA}
\affiliation[2]{Department of Astronomy, The Ohio State University, 4055 McPherson Laboratory, 140 W 18th Avenue, Columbus, OH 43210, USA}
\affiliation[3]{Department of Physics, The Ohio State University, 191 West Woodruff Avenue, Columbus, OH 43210, USA}
\affiliation{Remaining affiliations are in Appendix \ref{sec:affiliations}}
\emailAdd{karacayli.1@osu.edu}

\abstract{The one-dimensional power spectrum $P_{\mathrm{1D}}$ of Ly$\alpha$ forest offers rich insights into cosmological and astrophysical parameters, including constraints on the sum of neutrino masses, warm dark matter models, and the thermal state of the intergalactic medium. We present the measurement of $P_{\mathrm{1D}}$ using the optimal quadratic maximum likelihood estimator applied to over 300,000 Ly$\alpha$ quasars from Data Release 1 (DR1) of the Dark Energy Spectroscopic Instrument (DESI) survey. This sample represents the largest to date for $P_{\mathrm{1D}}$ measurements and is larger than the Extended Baryon Oscillation Spectroscopic Survey (eBOSS) by a factor of 1.7. We conduct a meticulous investigation of instrumental and analysis systematics and quantify their impact on $P_{\mathrm{1D}}$. This includes the development of a cross-exposure estimator that eliminates the need to model the pipeline noise and has strong potential for future $P_{\mathrm{1D}}$ measurements. We also present new insights into metal contamination through the 1D correlation function. Using a fitting function we measure the evolution of the Ly$\alpha$ forest bias with high precision: $b_F(z) = (-0.218\pm0.002)\times((1 + z) / 4)^{2.96\pm0.06}$.
In a companion validation paper, we substantially extend our previous suite of CCD image simulations to quantify the pipeline's exquisite performance accurately. In another companion paper, we present DR1 $P_{\mathrm{1D}}$ measurements using the Fast Fourier Transform (FFT) approach to power spectrum estimation. 
These two measurements produce a forest bias parameter that differs by 2.2 sigma. However, our model is simplistic, so this disagreement will be investigated in future work.
}

\begin{document}
\maketitle
\flushbottom


\section{Introduction\label{sec:intro}}
The Dark Energy Spectroscopic Instrument (DESI, \cite{leviDESIExperimentWhitepaper2013, desicollaborationDESIExperimentPart2016}) is on track to measure Baryon Acoustic Oscillations (BAO) with high precision across a broad range of redshifts. The DESI collaboration measured the BAO scale in 2024 \cite{desiKp6BaoLya2024, desiKp4BaoGalaxies2024, desiKp7Cosmology2024} from now-public Data Release 1 (DR1, \cite{desiKp2DataRelease12024}) and, most recently, in March 2025 from DR2 with over 14 million galaxies and 1.2 million quasars \cite{desiY3LyaBAO2025, desiY3BaoAndCosmology2025}. In the next years, DESI will collect data with accurate redshifts for at least 40 million galaxies and quasars over 14,000 square degrees.

At the highest redshifts, DESI probes cosmology with measurements of the \lya\ forest. Neutral hydrogen gas in the line-of-sight of distant quasars forms absorption lines at wavelengths shorter than the \lya\ emission line through absorption and scattering. These lines trace the underlying matter distribution in the intergalactic medium (IGM) and the circumgalactic medium (CGM). For a fiducial value of the sound horizon, the Lyman-$\alpha$ (\lya) forest in DR1 constrained the transverse comoving distance to $z_\mathrm{eff}=2.33$ with 2.4\% precision \cite{desiKp6BaoLya2024}, and with 1.3\% precision in DR2 \cite{desiY3LyaBAO2025}. However, the \lya\ forest data is full of valuable information at significantly smaller scales than the BAO scale of 150~Mpc through gas physics and structure formation.

The 1D power spectrum (\poned) is the key quantity to extract the physics down to scales below one Mpc from the \lya\ forest \cite{croftRecoveryPowerSpectrum1998, mcdonaldLyUpalphaForest2006, palanque-delabrouilleOnedimensionalLyalphaForest2013, irsicLymanEnsuremathAlpha2017, waltherNewPrecisionMeasurement2017, chabanierOnedimensionalPowerSpectrum2019, karacayliOptimal1DLy2022, karacayliOptimal1dDesiEdr2023, ravouxFFTP1dEDR2023, birdPriyaNewSuiteOfLyaSimulations2023, fernandezCosmologicaleBossPriya2024, waltherEmulatingLyaForestEboss2024}. \poned\ is especially sensitive to the thermal state of the IGM \cite{boeraRevealingThermal2019, waltherNewConstraintsIGM2019, villasenorThermalHistory2022}, and the amplitude and the slope of the linear matter power spectrum \cite{croftRecoveryPowerSpectrum1998, mcdonald_observed_2000, pedersenEmulator2021}. \poned\ has been used in a wide range of topics, including the primordial power spectrum \cite{vielPrimordialPowerSpectrumLya2004}, the sum of neutrino masses \cite{croftNeutrinoMassLyaForest1999, palanqueDelabrouilleNeutrinoMass2015, yecheNeutrinoMassesXQ2017},
and the nature of dark matter \cite{narayananWDMLyaForest2000, seljakSterileNeutrinosDM2006, wangLyaDecayingDM2013, irsicFuzzyDMfromLya2017, boyarskyLyaWDM2009, vielWarmDarkMatter2013, baurLyaCoolWDM2016, irsicConstraintsWDM2017, villasenorWarmDarkMatter2023}. Therefore, \poned\ is an important complementary measurement for DESI's mission. DESI will substantially expand the statistical power of the \lya\ forest measurements, requiring sophisticated and robust measurement tools of \poned\ and comprehensive studies of its systematics. 

In this paper, we measure the \lya\ forest \poned\ from DESI DR1 using the optimal estimator, also known as the quadratic maximum likelihood estimator (QMLE) \cite{mcdonaldLyUpalphaForest2006, karacayliOptimal1DLy2020, karacayliOptimal1dDesiEdr2023}. A companion paper presents the results from the FFT-based estimator \cite{ravouxFFTP1dDesiDr12024}, and another companion paper is dedicated to validating both \poned\ pipelines \cite{karacayliDesiY1P1dValidation}. The cosmological results are presented in \cite{chaves-monteroDesiDr1CosmologyP1d}.

The optimal estimator is fundamentally an inverse-covariance-weighted average ($\mathbf{C}^{-1}\bm{x}$). This foundation makes it ideally suited for \lya\ forest analyses, as it secures that measurements remain unaffected by gaps in data. Such gaps are ubiquitous in spectroscopy and especially in \lya\ forest data, where regions with strong atmospheric lines, high-column density (HCD) systems, and bad CCD pixels must be masked to mitigate contamination and instrumental imperfections. In the validation paper, we use 20 synthetic realizations of DESI DR1 to prove this property. Furthermore, QMLE can seamlessly integrate the individual resolution matrices of each spectrum, as provided by the DESI spectral extraction process, which is based on the spectro-perfectionism algorithm \cite{boltonSpectroPerfectionismAlgorithmicFramework2010, guySpectroscopicDataProcessingPipeline2022}. We validated this framework in our measurement from DESI's early data release (EDR, \cite{earlyDataRelease2023, surveyValidation2023}) by simulating CCD images of DESI quasar spectra and extracting them with the DESI spectroscopic pipeline \cite{karacayliOptimal1dDesiEdr2023}. The new validation paper substantially improves the statistical significance of this study, with a 15-fold increase in the number of simulations in order to more precisely quantify the residual biases in spectrograph resolution. 

Systematic error due to the pipeline noise estimation significantly degrades the constraining power of DESI \poned\ even though there are only percent-level calibration errors \cite{karacayliOptimal1dDesiEdr2023}. Quantifying the noise covariance matrix with the highest precision rapidly becomes impractical, despite our best attempts at granular calibration through data splits based on the spectrograph and CCD location of each quasar spectrum. However, a simple and robust way to avoid possible errors in the noise covariance matrix is to use cross-correlations among different exposures of the same quasar, following the common practice in CMB analyses \cite{hirataCmbLssWeakLensing2008, planckIntermediateResults2016, vannesteQuadaraticCmbCross2018}. In this cross-exposure QMLE (xQMLE) framework, systematic errors in different observations of the same quasar, such as noise, are independent. We develop and apply this estimator for the first time in a \lya\ forest \poned\ measurement. Unfortunately, nearly half of the DR1 quasar sample has only one exposure, so the results of this estimator mainly serve as a cross-check to the baseline estimator. However, this estimator will be extremely valuable for DR2, as most quasars have multiple exposures.

The outline of this paper is as follows. Section~\ref{sec:data} describes DESI DR1, the quasar catalog, and our refinements at high redshift, the catalog for quasars with broad absorption lines (BAL), and quasars with damped \lya\ absorption (DLA) systems. We explain our methodology in section~\ref{sec:methods}, including the quasar continuum fitting algorithm, QMLE and xQMLE methods, and the error estimation based on the bootstrap method. Section~\ref{sec:results} presents our main results. Here, we describe how we treat metal contamination and how we recalibrate the pipeline flux and variance estimates. In this section, we also present further details about the systematics that permeate our \poned\ measurement and the data and analysis variations we performed to corroborate our baseline results. We discuss our results, measure the forest bias, and provide recommendations for use in section~\ref{sec:discuss} and lastly summarize in section~\ref{sec:summary}. 

\section{Data\label{sec:data}}
Our sample is based on the quasar observations from DESI DR1 \cite{desiKp2DataRelease12024}), which is based on the first year of the main DESI survey, as well as earlier Survey Validation observations \cite{surveyValidation2023}. This sample has over 1.5 million quasars observed between December 2020 and June 2022 \cite{ashleyDesiLssCatalog2024}. Nearly 450,000 of these quasars are at $z > 2.1$; therefore, their \lya\ forest region falls onto DESI's wavelength coverage of 3600--9800~\AA. These spectra are collected with a new, high-throughput, multi-fiber spectrograph mounted on the 4-m Mayall telescope at Kitt Peak National Observatory in Arizona \cite{desicollaborationDESIExperimentPart2016b,abareshiOverviewInstrumentationDark2022}. Light enters the DESI instrumentation through a $3.2^\circ$ diameter wide-field corrector \cite{millerCorrector2024} that focuses the light onto a robotic focal plane assembly \cite{silberRoboticMultiobjectFocal2023} that was designed to obtain 5,000 spectra per observation with a reconfiguration time of under two minutes \cite{schlaflySurveyOps}. The light from the focal plane passes through a very high-efficiency fiber system \cite{poppettFiberSystem2024} to one of ten identical spectrographs that are in a climate-controlled room. The extraordinary stability of the spectrographs, combined with a regular calibration protocol and sophisticated data processing pipeline, all help to minimize instrumental systematic errors \cite{guySpectroscopicDataProcessingPipeline2022}.

\subsection{Quasar catalog}
DESI employs three automated classification algorithms to identify quasars and determine their redshifts \cite{chaussidonTargetSelectionDESIQSO2022}. \texttt{Redrock} is the main software package that determines the spectral class and redshift with a $\chi^2$ minimization analysis for a range of spectral templates based on principal component analysis (PCA) for each object \cite{brodzellerPerformanceOfQuasarTemplate2023}. Visual inspections during the survey validation period \cite{surveyValidation2023, alexanderDESISVVIQSO2022} demonstrated that \texttt{Redrock} misses some quasars. Therefore, DESI employs two additional afterburners to improve the quasar catalog: \texttt{QuasarNET}, which is a deep convolutional neural network \cite{buscaQuasarNET2018, farrQuasarNetDESI2020}, and an Mg~\textsc{ii} afterburner \cite{napolitanoMgIIAfterburner2023} that searches for broad Mg~\textsc{ii} emission without altering the redshift. As has been done in the Baryon Acoustic Oscillations (BAO) from the \lya\ forest analyses \cite{desiKp6BaoLya2024, desiY3LyaBAO2025}, we only keep quasars with \texttt{ZWARN=0} and \texttt{ZWARN=4}, eliminating quasars with any issue flagged by the spectroscopic pipeline.

\begin{figure}
    \centering
    \includegraphics[width=0.48\linewidth]{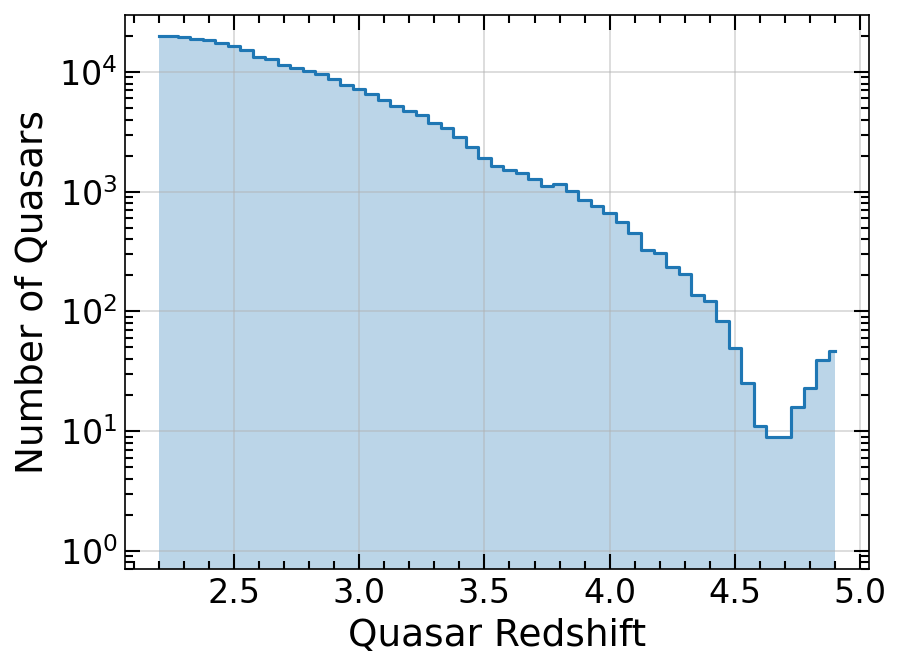} 
    \includegraphics[width=0.48\linewidth]{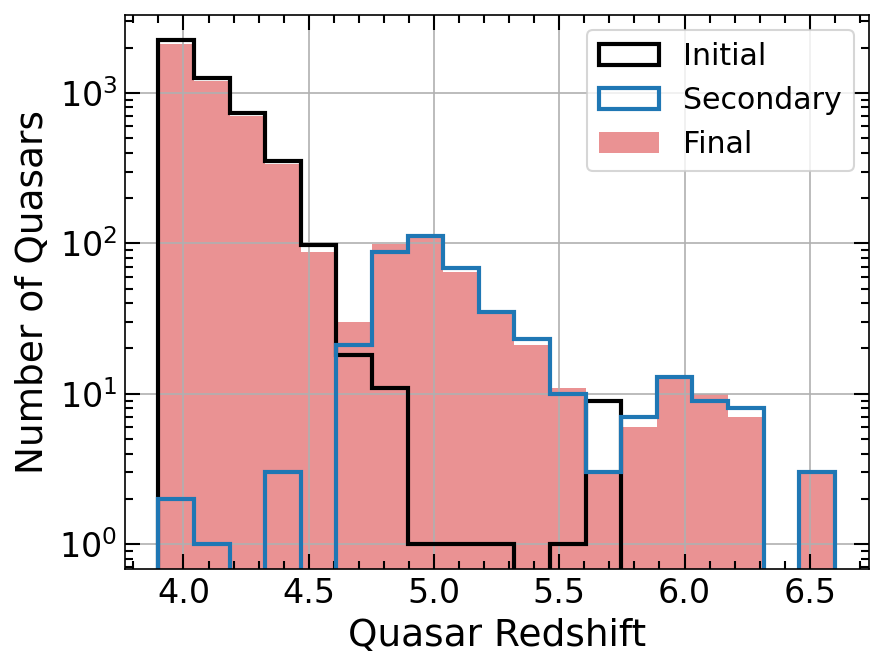}
    \caption{(Left) The redshift distribution of DESI DR1 quasar catalog after our high-redshift improvements. (Right) Change in the number of quasars at $z>3.9$. The black line is the initial quasar catalog. The blue line is the secondary program that we reintroduced to the catalog. The red-shaded region is the final quasar catalog after we eliminate the misclassified quasars identified through visual verification of 4,758 objects at $z>3.9$.}
    \label{fig:quasar_z_hist}
\end{figure}

We have performed two modifications to this base quasar catalog. First, we have conducted a visual inspection campaign for 4,758 objects at $z>3.9$ in the base catalog in order to ascertain the catalog's purity at higher redshifts. We found that 4,504 of these objects are high-redshift quasars, while the other 254 objects were misclassified. These numbers correspond to 94.7\% purity, a remarkable achievement of the automated pipeline in a difficult regime. However, stringent confidence cuts were needed to attain this high purity. Those cuts had removed 405 high-redshift quasars observed in a secondary program~\cite{yangDesiZ5QuasarSurvey2023}, and we have added those back to our quasar catalog. The left panel of figure~\ref{fig:quasar_z_hist} shows the redshift distribution of our final quasar catalog. The increase at $z>4.5$ is due to the reintroduction of the secondary program, which can be seen in the right panel that zooms in to $z>3.9$. These two refinements improve the quality of our measured \poned\ in the redshift range $3.6 < z <4.4$.

\subsection{Quasars with broad absorption lines}
Broad absorption line (BAL) features arise from quasar outflows and can impact quasar redshift estimates and classifications, and contaminate the \lya\ forest. These features are identified in DR1 using the same approach described in ref.~\cite{filbertBALEDRcatalog2023}, which was used in the DESI early data \poned\ analyses~\cite{karacayliOptimal1dDesiEdr2023, ravouxFFTP1dEDR2023}. We find that 18.5\% of our DR1 quasars at $z>2.1$ have BAL features. 
The BAL identification method measures a velocity range for each trough associated with the C~\textsc{iv} emission feature at 1549~\AA. As an improvement over the EDR analysis, we mask the wavelength ranges of the absorption troughs of S~\textsc{iv}, P~\textsc{v}, C~\textsc{iii}, C~\textsc{iv}, \lya, N~\textsc{v}, and Si~\textsc{iv} that correspond to the same velocities. Keeping quasars with BAL features while masking the associated metal lines recovers high SNR spectra in which these systems are identified and boosts the constraining power of our \poned\ measurement. This has been shown to give unbiased results in the validation paper \cite{karacayliDesiY1P1dValidation}. Furthermore, we confirm in Section~\ref{subsec:highsnr} that removing BAL quasars identified by the balnicity index while masking the features associated with the absorption index --- which is the same strategy used in the FFT paper \cite{ravouxFFTP1dDesiDr12024} --- gives effectively the same results.

\subsection{Damped \texorpdfstring{\lya}{Lya} absorption systems (DLAs)\label{subsec:data_dla_catalog}}
The models for the \lya\ forest \poned\ assume the density fluctuations are in the quasi-linear regime. Damped \lya\ systems with high column densities ($N_\mathrm{HI}>2 \times 10^{20}$~cm$^{-2}$) are complex to model in hydrodynamical numerical simulations fully. When unmasked, the extensive damping wings of these systems add power to the largest scales.
While empirical models of DLAs can be constructed and are useful in marginalizing some of their contribution to \poned\
\cite{mcdonaldPhysicalDampingWings2005, font-riberaEffectHighColumn2012, rogersCorrelations3dHighColumnDensity2018, rogersEffectsOfHcd2018}, these models cannot fully account for the clustering strength of DLAs with the underlying matter field and are coarsely categorized in terms of column densities. 

Therefore, it has been the norm in the literature to mask the highest column density systems, which have the largest impact, as best as possible, and tackle the incompleteness of the masking process by assigning a systematic error budget to the measurement or by using the templates mentioned above in cosmological inference \cite{palanque-delabrouilleOnedimensionalLyalphaForest2013, chabanierOnedimensionalPowerSpectrum2019, karacayliOptimal1dDesiEdr2023, ravouxFFTP1dEDR2023}. In this analysis, we identify DLAs using two separate identifiers and combine the two catalogs into a ``concordance" catalog. The Convolutional Neural Network (CNN) identifier and its performance in simulated DESI spectra are presented in ref.~\cite{wangDeepLearningDESIDLA2022}. The identifier based on the Gaussian process (GP) model is presented in ref.~\cite{mingfengDLAGP2021}. 
GP results are adopted in the concordance catalog over CNN results for systems that are detected by both.
The need to remove as many of these systems as possible is significantly larger than our companion BAO analysis, since they do not bias the BAO location as they are another correlated tracer, but will bias cosmological results obtained from \poned. Therefore, we opt for looser confidence cuts in DLA identification than the ones used in ref.~\cite{desiKp6BaoLya2024}. To assess the effectiveness of our catalog, we perform variations in confidence cuts to create ``high-confidence" and ``high-completeness" catalogs. We provide the details in Appendix~\ref{app:dla_cat}. In short, the high-completeness catalog has 122,458 DLAs in 85,186 quasars, whereas the high-confidence catalog has 83,795 DLAs in 57,909 quasars. They both have the same average of 1.4 DLAs per sightline that has at least one DLA. Even though the high-confidence catalog has 38,663 fewer DLAs, we do not expect its purity to be substantially better than the high-completeness catalog \cite{brodzellerConstructionDlaDesiDr2}. We will discuss implications of this in section~\ref{subsec:dla_syst}.

\section{Methods\label{sec:methods}}

\subsection{Continuum fitting}
We perform quasar continuum fitting with the same algorithm as in DESI EDR \cite{bourbouxCompletedSDSSIVExtended2020, karacayliOptimal1dDesiEdr2023, ravouxFFTP1dEDR2023}, although with a different software package called \qsonic\footnote{\url{https://qsonic.readthedocs.io/en/stable/}}\cite{Karacayli_QSOnic_fast_quasar_2024}. We provide a short overview here, and refer the reader to refs.~\cite{bourbouxCompletedSDSSIVExtended2020, ramirezperezLyaCatalogDesiEdr2023, karacayliOptimal1dDesiEdr2023} for more details.

The algorithm assumes every quasar continuum $C_q$ can be described by a mean quasar continuum $\overline{C}(\lambda_\mathrm{RF})$, where $\lambda_\mathrm{RF}$ is the wavelength in quasar's rest frame, plus two quasar-specific terms: an amplitude $a_q$ and a slope $b_q$, such that $C_q (\lambda_\mathrm{RF}) = ( a_q + b_q \Lambda) \overline{C}\left(\lambda_\mathrm{RF} \right)$, where $\Lambda \sim \log\lambda_\mathrm{RF}$.
The \qsonic\ package multiplies this continuum expression with an input mean IGM transmission $\overline{F}(z)$ to obtain $\delta_F = f/\overline{F}C_q - 1$, the transmitted flux fluctuations. It then iteratively refines the mean quasar continuum estimate. The $a_q$ and $b_q$ values are calculated by minimizing the following function:
\begin{equation}
    \chi^2 = \sum_j \frac{\left[f_j - \overline{F}(z_j) ( a_q + b_q \Lambda_j) \overline{C}\left(\frac{\lambda_j}{1 + z_q}\right) \right]^2}{\sigma_{q, j}^2} + \sum_j \ln \sigma_{q, j}^2,
\end{equation}
where the summation $j$ is over all pixels in the forest region, $\lambda_j$ is the observed wavelength, $z_j = \lambda_j / \lambda_\mathrm{Ly\alpha} - 1$ and $\Lambda_j \sim \log\lambda_\mathrm{RF, j}$.  These $a_q, b_q$ parameters also absorb the effect of the mean IGM flux and the underlying large-scale overdensity field, which biases the quasar continuum and resulting \poned. This is addressed in section~\ref{subsubsec:continuum}. The variance of each pixel $\sigma_{q, j}^2$ is the sum of the pipeline variance $\sigma^2_\mathrm{pipe}$ and the large-scale \lya\ fluctuations $\sigma^2_\mathrm{LSS}$:
\begin{equation}
    \sigma_{q, j}^2 = \eta(\lambda_j) \sigma^2_\mathrm{pipe, j} + \sigma^2_\mathrm{LSS}(\lambda_j) ( a_q + b_q \Lambda_j)^2 \overline{F}^2(z_j) \overline{C}^2\left(\frac{\lambda_j}{1 + z_q}\right) \label{eq:sigma2_cfit},
\end{equation}
where $\eta(\lambda)$ is the pipeline noise correction term. After every quasar is fit, we stack all continua in the rest frame, update the mean continuum  $\overline{C}$, and calculate $\eta$ and $\sigma^2_\mathrm{LSS}$ as described in the references. The mean continuum is computed using coarse rest-frame binning pixels with $\Delta \lambda_\mathrm{RF}=0.4~$\AA\ and converges after five iterations.

We mask DLAs and BALs before the continuum fitting. For each DLA in a quasar spectrum, we calculate its transmission profile using a Voigt profile and mask regions with more than 20\% absorption while correcting the remaining wings. All possible absorption regions associated with BAL quasars are masked without correction. We also masked major atmospheric lines\footnote{\url{https://github.com/corentinravoux/p1desi/blob/main/etc/skylines/list_mask_p1d_DESI_EDR.txt}}.

\subsection{Optimal estimator}
The quadratic maximum likelihood estimator (QMLE), also known as the optimal estimator, operates in real space, which facilitates individually weighting pixels and accounting for the intrinsic \lya\ large-scale structure correlations in the weights. However, the most important property of QMLE is that it is not biased by gaps in the spectra, unlike FFT-based estimators. In other words, QMLE deconvolves and eliminates the ``survey function" from biasing the measurement and subsequent inferences. The QMLE formalism has been extremely successful in applications to the cosmic microwave background radiation, galaxy surveys, and weak lensing \cite{hamiltonOptimalMeasurementPower1997, tegmarkKarhunenLoeveEigenvalueProblems1997, tegmarkMeasuringGalaxyPower1998, seljakWeakLensingReconstruction1998}. In the \lya\ \poned\ literature, it has been applied to a variety of data sets such as SDSS \cite{mcdonaldLyUpalphaForest2006}, high-resolution, high-SNR spectra \cite{karacayliOptimal1DLy2022}, and DESI early data \cite{karacayliOptimal1dDesiEdr2023}. We refer the reader to refs.~\cite{karacayliOptimal1DLy2020, karacayliOptimal1DLy2022} for our development process and implementation details, and to ref.~\cite{karacayliDesiY1P1dValidation} for a description of the extensive performance updates to our code since DESI's early data analysis. The software package is called \texttt{lyspeq}\footnote{\url{https://github.com/p-slash/lyspeq}}.

Briefly, QMLE gives a power spectrum $\bm{p}_q$, and its Fisher (inverse covariance\footnote{The Fisher matrix is sometimes referred to as the mode-mixing matrix.}) matrix $\mathbf{F}_q$ for each quasar $q$, and the final product is a weighted average: $\bm{p} = \mathbf{C}_\mathrm{tot} \sum_q \mathbf{F}_q \bm{p}_q$, where $\mathbf{C}_\mathrm{tot}^{-1} \equiv \sum_q \mathbf{F}_q$. The total covariance matrix $\mathbf{C}_\mathrm{tot}$ normalizes the power spectrum based on propagated pixel weights and also deconvolves the survey window function that mixes different $(k, z)$ modes due to masking, continuum marginalization, and reverse interpolation of pixel pairs to two redshift bins.

Our QMLE implementation estimates deviations from a fiducial power spectrum $P_{\mathrm{fid}}(k, z)$, which is also employed for inverse covariance weighting of each spectrum:
\begin{equation}
    \label{eq:pd13_fitting_fn}\frac{kP_{\mathrm{fid}}(k, z)}{\pi} = A \frac{(k/k_0)^{3 +n + \alpha\ln k/k_0}}{1+(k/k_1)^2} \left(\frac{1+z}{1+z_{0}}\right)^{B + \beta\ln k/k_0},
\end{equation}
where $k_{0} = 0.009~$\skm\ and $z_{0}=3.0$. Estimating the deviations from the fiducial mitigates the effects of bin averaging within $k$ bins \cite{karacayliOptimal1DLy2020}.

We perform a single iteration to estimate the power spectrum since further iterations refine the Fisher matrix estimate \cite{karacayliOptimal1DLy2020}, rather than improve the power spectrum estimate. The QMLE's Fisher (covariance) matrix does not constitute a sufficiently accurate error estimate, so these improvements are not worth the computing time. We instead estimate the covariance of our measurement using a regularized bootstrap method described below. Spectra with more than 350 pixels are split into two segments to reduce the computation time of matrix formation and multiplication (dominated by Fisher matrix computation) and help continuum marginalization. Segments with fewer than 40 pixels are discarded from the estimation.

\subsubsection{Cross-exposure estimator}
Estimating the power spectrum from auto-correlations of coadded spectra (auto-spectra) is the most powerful method to yield the lowest uncertainties on the measured \poned. However, auto-spectra estimators of \poned\ come with added spectral noise power, which must be subtracted from the \poned\ estimates. This is the conventional method in the literature, including this work. As the uncertainties shrink, the accuracy of this noise power becomes a significant factor in the precision of \poned\ estimates and a source of substantial systematics. This noise power can be eliminated when correlations between independent measurements are used to estimate \poned, as has been the norm in CMB measurements \cite{hirataCmbLssWeakLensing2008, planckIntermediateResults2016, vannesteQuadaraticCmbCross2018}. Instead of using coadded spectra, we build an estimator of \poned\ that cross-correlates multiple exposures of the same quasar and call this estimator the cross-exposure estimator.

The QMLE of auto-spectra is an optimal estimator that achieves the minimum variance for Gaussian fields. This formalism can be extended to construct an unbiased cross-exposure estimator that preserves inverse-covariance weighting and remains resilient to missing data. However, while this cross-exposure estimator retains these advantages, it does not inherit the theoretical minimum variance property of the auto-spectra estimator \cite{planckIntermediateResults2016}. In the cross-exposure QMLE (xQMLE), each exposure $m$ is weighted by its inverse covariance $\bm{y}_m = \mathbf{C}^{-1}_m \bm{\delta}_m$, then cross-correlated with another exposure $n$ using the derivative matrix $\mathbf{C}_{,k}$ for a $k$ bin. The xQMLE is formulated similarly to the auto-spectra estimator without the noise power term ($b_k$ in ref.~\cite{karacayliOptimal1DLy2020}):
\begin{equation}
    \hat P_k = P_\mathrm{fid}(k) + \frac{1}{2} \sum_{k'} F^{-1}_{kk'} (d_{k'} - t_{k'}),
\end{equation}
where
\begin{gather}
    d_k = \sum_m \sum_{n\neq m} \bm{y}_m^T \mathbf{C}^{(mn)}_{,k} \bm{y}_n, \qquad
    t_k = \sum_m \sum_{n\neq m} \Tr\left(\mathbf{C}^{-1}_m \mathbf{C}^{(mn)}_{,k} \mathbf{C}^{-1}_n \mathbf{S}^{(nm)}_{\mathrm{fid}}\right), \\
    F_{kk'} = \sum_m \sum_{n\neq m} \frac{1}{2} \Tr\left( \mathbf{C}^{-1}_m \mathbf{C}^{(mn)}_{,k} \mathbf{C}^{-1}_n \mathbf{C}^{(nm)}_{,k'} \right),
\end{gather}
and $\mathbf{S}_{\mathrm{fid}}$ is the signal contribution to the covariance matrix based on $P_\mathrm{fid}(k)$.
As noted in the literature, the ``Gaussian" covariance matrix $\mathbf{F}^{-1}$ is not accurate as an error estimate even though it is the unbiased window function for the cross-exposure estimator \cite{planckIntermediateResults2016, vannesteQuadaraticCmbCross2018}.

\subsubsection{Regularized bootstrap covariance matrix}
The non-parametric bootstrap method \cite{efronBootstrap1979} is efficient in capturing non-Gaussian contributions to the covariance matrix due to non-linearities in the \lya\ forest and other effects in instrumentation, such as Poisson noise. This method typically requires $\mathcal{O}(N^2)$ independent data samples to reduce the random fluctuations in the estimated $N\times N$ covariance matrix. In the DESI EDR analysis, \texttt{lyspeq} was limited to using MPI tasks ($N_\mathrm{task}=256$) as subsamples, which yielded largely noisy covariance matrix estimates since the entire $N$-dimensional vector space could not be explored. \texttt{lyspeq} can now bootstrap over quasars with a simplifying assumption that the Fisher matrix remains the same for all bootstrap realizations (i.e., the Fisher matrix is not recalculated for every realization). Second, it applies a Poisson bootstrap method, where observation frequencies for each quasar are drawn from a Poisson distribution with a mean of one\footnote{This means we do not have the same number of quasars across bootstrap realizations.} \cite{hanleyNonParametricPoissonBootstrap2006, chamandy2012estimating}. These are reasonable approximations since the sample size is large.

Unfortunately, the estimated covariance matrix still has noise problems, mostly at higher redshifts where there are only a few hundred quasars per bin. Our solution is to ``regularize" this matrix by smoothing the covariance matrix and enforcing that the final bootstrap covariance matrix is larger than the Gaussian (optimal estimator) covariance matrix $\mathbf{C}_\mathrm{B} \geq \mathbf{C}_\mathrm{G}$.
Here is a summary of the key steps. First, we normalize the Gaussian and bootstrap matrices with the diagonals of the Gaussian matrix. The difference between these matrices is smoothed with a 2D Gaussian filter across rows and columns with $\sigma=2$. The smoothed bootstrap estimates $\widetilde{\mathbf{C}}_\mathrm{B}^{(1)}$ are constructed by summing this matrix and the Gaussian matrix. Since this is not guaranteed to be larger than the Gaussian matrix, we first follow refs.~\cite{mcdonaldLyUpalphaForest2006, karacayliOptimal1dDesiEdr2023} and replace the eigenvalues $\lambda_i\rightarrow\mathrm{max}(\lambda_i, \bm e_i^\mathrm{T} \mathbf{C}_\mathrm{G} \bm e_i)$, where $\bm e_i$ are the eigenvectors of $\widetilde{\mathbf{C}}_\mathrm{B}^{(1)}$, to obtain an intermediate matrix $\widetilde{\mathbf{C}}_\mathrm{B}^{(2)}$. The final bootstrap covariance is formulated as $\widetilde{\mathbf{C}}_\mathrm{B}^\mathrm{final} = \mathbf{C}_\mathrm{G} + \mathbf{M}^+$, where $\mathbf{M}^+$ is calculated by setting all negative eigenvalues of $\mathbf{M} = \widetilde{\mathbf{C}}_\mathrm{B}^{(2)} - \mathbf{C}_\mathrm{G}$ to zero. This process is described in more detail in the companion validation paper \cite{karacayliDesiY1P1dValidation}.

\section{Results\label{sec:results}}

\subsection{Baseline and cross-exposure measurements}
We restrict the \lya\ forest region to the conservative rest-frame range of 1050--1180~\AA\  of each quasar, following the standard in the literature, and to the observed wavelength range of 3600--7000~\AA, based on the coverage of our data. Forests that retain less than 20\% of this rest-frame wavelength range after masking BAL features, DLAs, and sky emission lines are excluded from the analysis. Additionally, to enhance the purity of the quasar, DLA, and BAL catalogs, we remove forests with a mean SNR below 0.3 per pixel in the forest region or below one on the red side of the \lya\ emission line, $\lambda>(1 + z_\mathrm{q})\lambda_\mathrm{Ly\alpha}$.

The noise and flux reported by the pipeline contain calibration errors that require correction. As detailed below, we derive calibration parameters using sidebands (SBs) at longer wavelengths than the \lya\ emission line in the quasar's rest frame. Neutral hydrogen cannot absorb radiation emitted at these wavelengths, so only metals with $\lambda\gtrsim 1300$~\AA\ transition wavelengths form (typically weak) absorption lines,
resulting in a smoother transmission field in SBs. This property makes SBs better-suited for calibration purposes as opposed to the heavily absorbed and correlated \lya\ forest region. Since the same metal lines contaminate the forest, these SBs are also utilized to statistically estimate metal contamination in \poned\ \cite{mcdonaldLyUpalphaForest2006, palanque-delabrouilleOnedimensionalLyalphaForest2013, chabanierOnedimensionalPowerSpectrum2019, karacayliOptimal1dDesiEdr2023}. This metal power spectrum is subsequently subtracted from our final measurements.
Table~\ref{tab:qsonumbers} lists the wavelength ranges for \lya\ and SB regions and the number of quasars in each region. 
\begin{table}
    \centering
    \begin{tabular}{|l|c|c|}
        \hline
        Region & Wavelength range [\AA] & Number of quasars \\
        \hline
        \lya\ & 1050--1180 & 314,241 \\
        SB1 & 1268--1380 & 612,904 \\
        SB2 & 1409--1523 & 786,591 \\
        SB3 & 1600--1800 & 961,691 \\
        \hline
    \end{tabular}
    \caption{Rest-frame wavelength ranges and number of quasars in DESI DR1.}
    \label{tab:qsonumbers}
\end{table}
We use the power spectrum estimated in SB1 $(P_\mathrm{SB1})$ as the metal power spectrum. Subtracting $P_\mathrm{SB1}$ removes power contributions from most metals $(\lambda\gtrsim 1300~\text{\AA})$. However, some residual metal contamination remains, such as Si~\textsc{iii}~1207~\AA, which introduces oscillatory features.

We estimate \poned\ in 60 linear bins with $\Delta k_\mathrm{lin}=5\times10^{-4}~$\skm and 25 log-linear bins with $\Delta k_\mathrm{log}=0.01$, and in redshift bins of size $\Delta z=0.2$ from $z=2.0$ to $z=4.6$. We remove the first and last redshift bins from our estimate since these are susceptible to edge effects. Furthermore, based on our validation tests \cite{karacayliDesiY1P1dValidation}, we recommend the following $k$ cuts: $k>10^{-3}~$\skm\ due to continuum error contamination and $k < 0.5 \pi / R_z$, where $R_z \equiv c \Delta\lambda_\mathrm{DESI} / (1 + z) \lambda_\mathrm{Ly\alpha}$ and $\Delta\lambda_\mathrm{DESI}=0.8$~\AA, based on a conservative estimate of the size of the spectrograph resolution correction. We calculate the fiducial power in eq.~\eqref{eq:pd13_fitting_fn} with the best-fit values from the DESI EDR measurement in the \lya\ forest region: $A=0.0763, n=-2.52, \alpha=-0.128, B=3.67, \beta=-0.286$ and $k_1=0.0369$\,s\,km$^{-1}$.

We apply the cross-exposure estimator (xQMLE) with the same parameters as the baseline (auto-spectra) estimator. Specifically, we cross-correlate exposures of the same quasar taken on different nights and with different spectrographs in order to mitigate potential correlations arising from sky modeling and CCD master biases. At $z>3.8$, the reduced number of exposures results in increased noise and inaccuracy. So these additional redshift bins are removed from the xQMLE calculation.

\begin{figure}
    \centering
    \includegraphics[width=\linewidth]{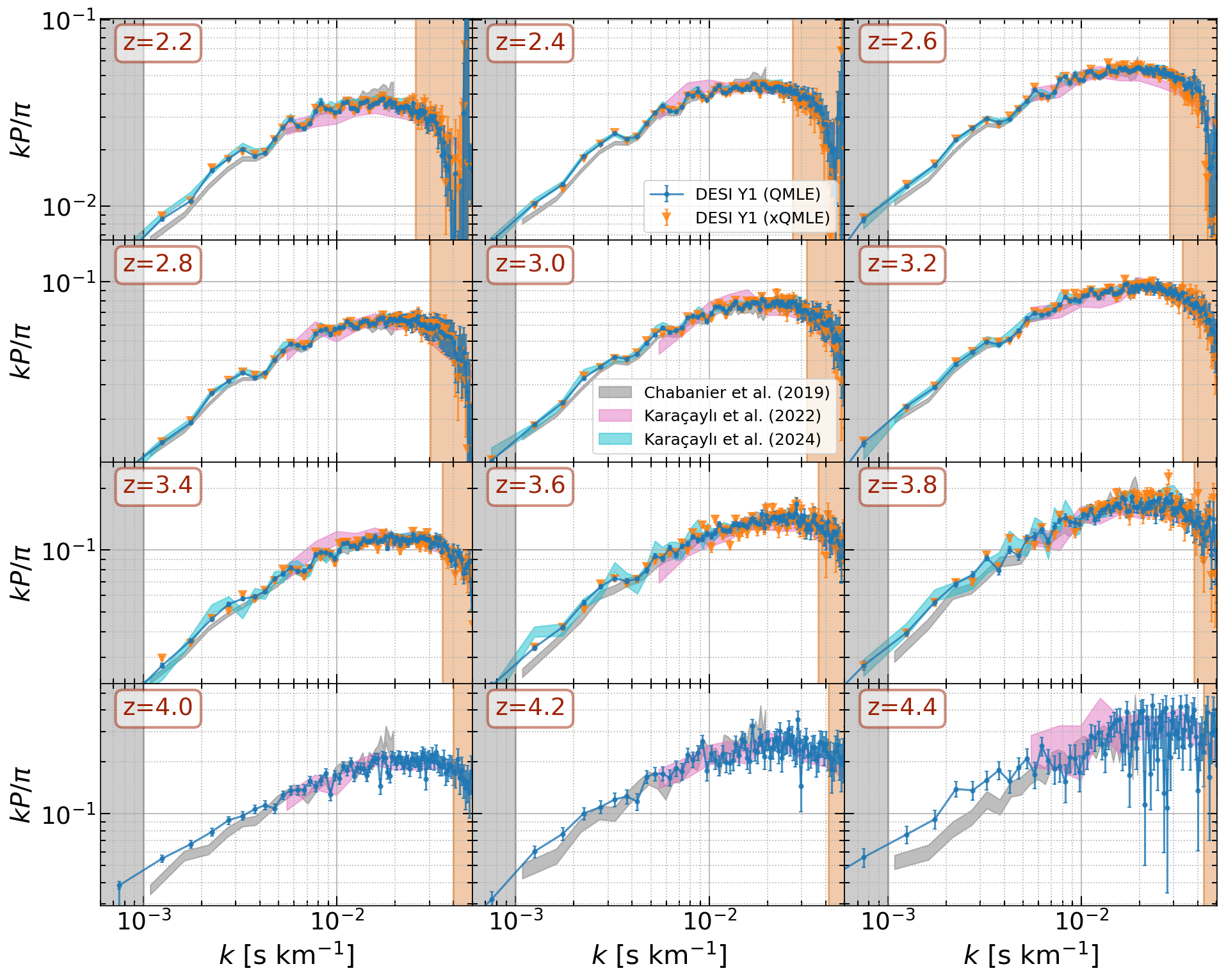}
    \caption{DESI DR1 baseline QMLE results ({\it blue circles with solid line}) and cross-exposure xQMLE results ({\it orange triangles}). The metal contamination is subtracted from both estimates. The error bars include systematic errors. xQMLE is shown until $z=3.8$ since having fewer exposures at higher redshifts makes the estimates noisier. Both methods give consistent results. However, xQMLE shows more scatter at $z=3.6$ and $z=3.8$. Though the small sample size could cause these fluctuations, the transition from blue to red channel also occurs between these two redshifts. Our findings remain consistent with DESI EDR results ({\it shaded light blue}) and are in 15\% tension with eBOSS results ({\it shaded grey}). The \poned\ measured from high-resolution spectra of KODIAQ, SQUAD, and XQ-100 ({\it shaded pink}) displays a broad agreement with our results.}
    \label{fig:qmle_y1_p1d}
\end{figure}

Our baseline QMLE and xQMLE \poned\ results from DESI DR1 are presented in figure~\ref{fig:qmle_y1_p1d}.
The estimates from these two methods are consistent and also align with DESI EDR results \cite{karacayliOptimal1dDesiEdr2023} shown in shaded light blue; however, they remain $5-15\%$ larger than the eBOSS measurements \cite{chabanierOnedimensionalPowerSpectrum2019}. The most probable explanation for this discrepancy is the HCD contamination (though see section~\ref{sec:discuss} for details). We test for impurity and incompleteness by varying our DLA catalog and find no evidence of errors. However, this is a correlated error mode, so coherent fluctuations in either measurement could account for the discrepancy. Our measurement is also broadly consistent with the \poned\ measured from the combined data set of KODIAQ \cite{omearaSecondDataRelease2017}, SQUAD \cite{murphyUVESSpectralQuasar2019}, and XQ-100 \cite{lopezXQ100LegacySurvey2016} high-resolution spectra \cite{karacayliOptimal1DLy2022}. The scatter in xQMLE results at $z=3.6$ and $z=3.8$ exceeds the estimated error bars. While this could be attributed to the small sample size leading to statistical fluctuations and underestimated uncertainties, it also coincides with the transition from the blue to the red channel, which may contribute to the observed fluctuations.

\subsection{Metal contamination from side bands}
Wavelengths longer than the \lya\ emission line in quasar spectra are populated by metal absorption lines but are free from neutral hydrogen absorption. We use these wavelengths to statistically estimate metal contamination in the \lya\ forest \cite{mcdonaldLyUpalphaForest2006, palanque-delabrouilleOnedimensionalLyalphaForest2013, chabanierOnedimensionalPowerSpectrum2019, karacayliOptimal1dDesiEdr2023, ravouxFFTP1dEDR2023} and to correct (recalibrate) the pipeline flux and noise calibration errors \cite{karacayliOptimal1dDesiEdr2023}. This correction is described in detail in section~\ref{subsec:flux_noise_calib}. In all cases where we use the SBs, the continuum fitting follows the same procedure as in the \lya\ forest region, except that no DLA masking is applied since DLAs do not exist in the SBs. Additionally, the mean transmission in \qsonic\ is set to one for all redshifts. The following fiducial power spectrum parameters are used in the covariance matrix calculation: $A=0.00208$, $n=-3.075$, $\alpha=-0.0742$, $B=1.599$, $\beta=-0.238$, and the $k_1$ term is ignored. We estimate \poned\ directly rather than its deviations from $P_\mathrm{fid}$. This is because eq.~\eqref{eq:pd13_fitting_fn} is not as good a description of the metal power spectrum as \lya\ as it lacks strong metal doublet transitions such as C~\textsc{iv}.

We obtain the final metal power spectrum after the flux and noise recalibration in SB1, which includes the most metal lines among the three SBs. This is verified by the fact that $P_\mathrm{SB1} > P_\mathrm{SB2}$, which can be seen in the right panel of figure~\ref{fig:calib_eta_flux}. As refs.~\cite{karacayliOptimal1dDesiEdr2023, karacayliFrameworkMetals2023} showed, the C~\textsc{iv} doublet is the dominant feature, which manifests as an oscillation in the power spectrum, while other doublets appear as smaller oscillations enveloped by this prominent signal. The statistical uncertainties in this measurement are propagated to our final SB1 subtracted measurement: $\mathbf{C}_\mathrm{final} = \mathbf{C}_\mathrm{Ly\alpha} + \mathbf{C}_\mathrm{SB1}$.

\subsection{Correcting pipeline flux and noise calibration\label{subsec:flux_noise_calib}}
The DESI pipeline flux and noise (variance) estimates are affected by percent-level calibration errors \cite{karacayliOptimal1dDesiEdr2023, guyCharacterizationOfContaminants2025}. Errors in noise estimation directly propagate to the final \poned\ measurements in auto-spectra estimators, whereas cross-exposure estimates remain robust against pipeline noise errors by construction. SBs provide smooth regions in quasar spectra that allow for calibration of the pipeline variance by comparing it to the observed variance. Since SBs can be observed in lower-redshift quasars, their sample size is significantly larger than that of the \lya\ forest data alone, which improves the statistical reliability.

\begin{figure}
    \centering
    \includegraphics[width=0.58\linewidth]{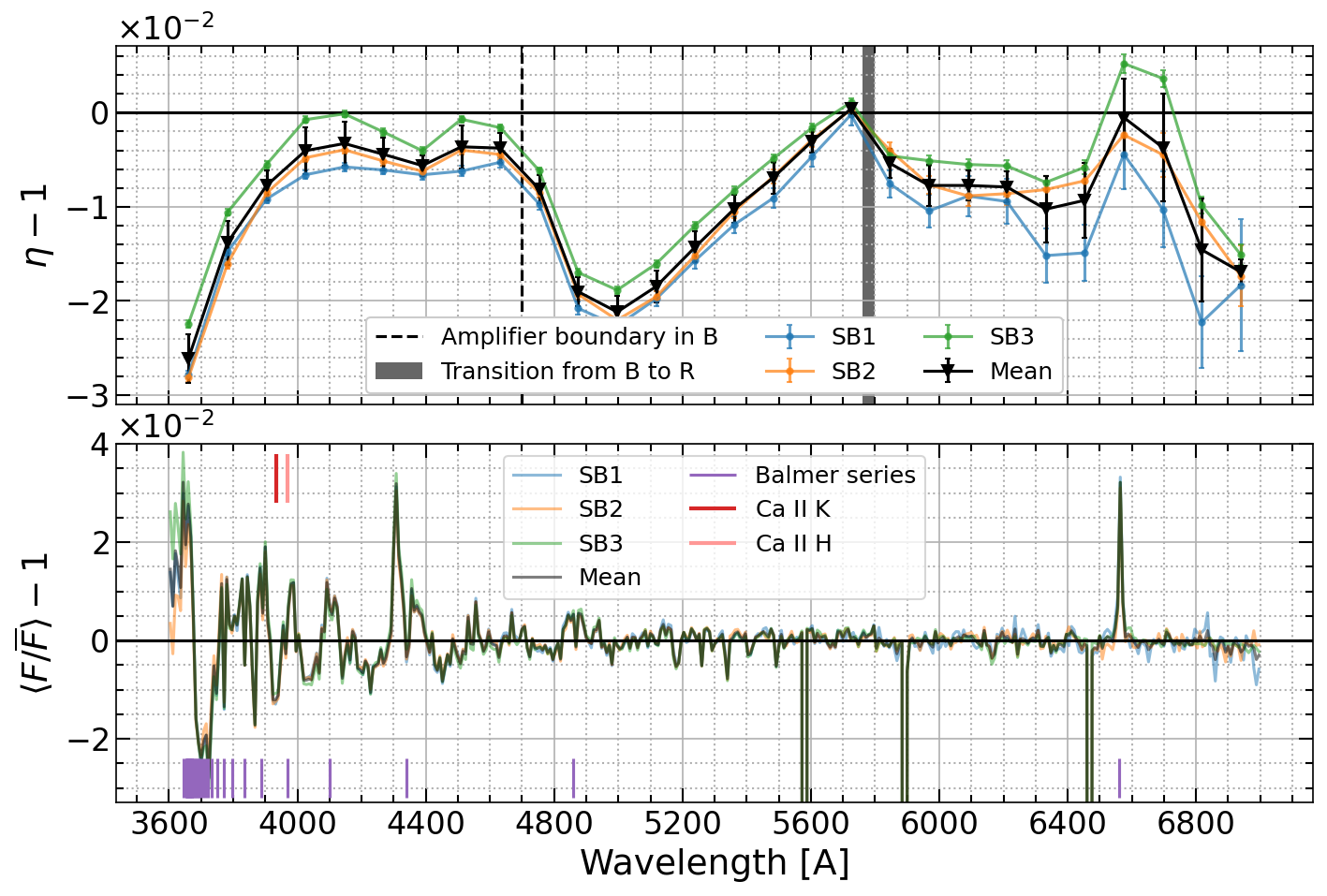}
    \hfill
    \includegraphics[width=0.4\linewidth]{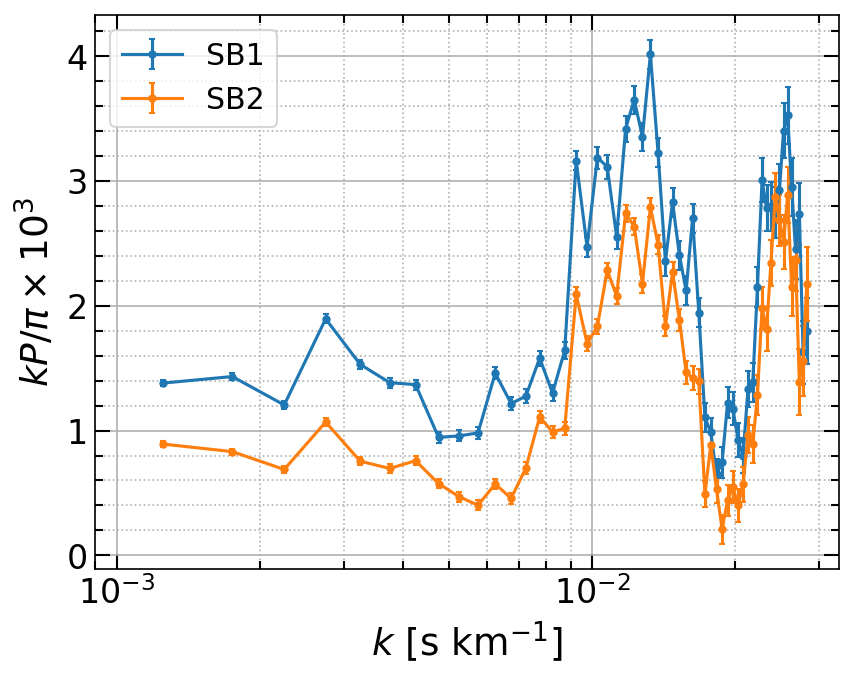}
    \caption{({\it Top left}) Pipeline noise (variance) correction factors $\eta$ across all three SBs exhibit similar percent-level features. Sharp features occur near the transition between CCD amplifiers in the blue channel ({\it dashed line}) and in the overlap region between the blue and red channels ({\it shaded}). The average $\overline{\eta}$ ({\it black triangles}) is our final pipeline noise correction factor. ({\it Bottom left}) The stack of $F/\overline{F} - 1 = \delta_F$ in the observed wavelength frame across all three SBs. The largest flux calibration errors, reaching up to 4\%, are observed at the locations of the Balmer series and Ca~\textsc{ii} H\&K doublet lines. The mean ({\it black line}) is used to recalibrate the spectra. ({\it Right}) The power spectra estimated from two SBs in $z=2.6$. SB1 includes the most metal lines as verified by $P_\mathrm{SB1} > P_\mathrm{SB2}$. The C~\textsc{iv} doublet is the dominant feature, manifesting as an oscillation in the power spectrum.
    }
    \label{fig:calib_eta_flux}
\end{figure}

The noise calibration errors are represented by the $\eta$ parameter in the continuum fitting algorithm \cite{bourbouxCompletedSDSSIVExtended2020, ramirezperezLyaCatalogDesiEdr2023, karacayliOptimal1dDesiEdr2023}. The \qsonic\ package estimates this parameter by first calculating the $\langle\delta_F\rangle, \langle\delta_F^2\rangle, \langle\delta_F^4\rangle$ moments in logarithmic $\sigma^2_\mathrm{pipe}$ bins for each redshift bin. The covariance of each moment is estimated using a delete-one Jackknife method over sub-samples. The $\eta$ parameter in each redshift bin is fitted using the total covariance matrix of $\sigma^2_\mathrm{obs} = \langle\delta_F^2\rangle - \langle\delta_F\rangle^2$ \cite{Karacayli_QSOnic_fast_quasar_2024}. Like the EDR analysis, all three SBs yield similar $\eta$ values, as shown in the top left panel of figure~\ref{fig:calib_eta_flux}. The boundary between CCD amplifiers in the blue channel is approximately indicated by the black dashed line at 4700~\AA, while the overlap region between the blue and red channels is shaded in black at 5780~\AA. Sharp features occur near these transition regions. The average correction factor for pipeline noise, $\overline{\eta}$, is represented by black triangles.

We estimate the flux calibration errors by stacking the SB transmission in the observed wavelength frame on a grid of 8~\AA, which smoothes out fluctuations below this scale that would be considered spurious for flux calibration purposes. The bottom left panel of figure~\ref{fig:calib_eta_flux} shows the residuals across all three SBs and their mean. The largest flux calibration errors are observed at the locations of the Balmer series and Ca~\textsc{ii} H\&K doublet lines. The mean of $\langle F/\overline{F} \rangle - 1$ across all three SBs represents the flux calibration correction factor used in this analysis and is shown by the black line.

\subsection{Systematics}
There are five sources of major systematic errors in DESI DR1 \poned\ for the quadratic estimator. First, our DLA catalog's completeness and purity are around 85\% \cite{wangDeepLearningDESIDLA2022}, so missing DLAs and false positives would directly contaminate \poned. Similarly, our BAL catalog is incomplete, and the effect of the remaining BAL features must be considered in the error budget.
Third, our continuum fitting algorithm effectively removes underlying large-scale modes, biasing the \poned\ mostly on large scales but also across all scales due to higher-order correlations that contaminate the two-point statistics. Fourth, the spectrograph resolution restricts the smallest scale that DESI can conservatively measure to its half-Nyquist frequency and adds significant uncertainties at small scales to account for its uncertainty in deconvolution. Finally, the pipeline noise calibration errors and correlated CCD read noise are critical sources of uncertainty since the noise power spectrum is subtracted to form the baseline measurement. 

\begin{figure}
    \centering
    \includegraphics[width=1.0\linewidth]{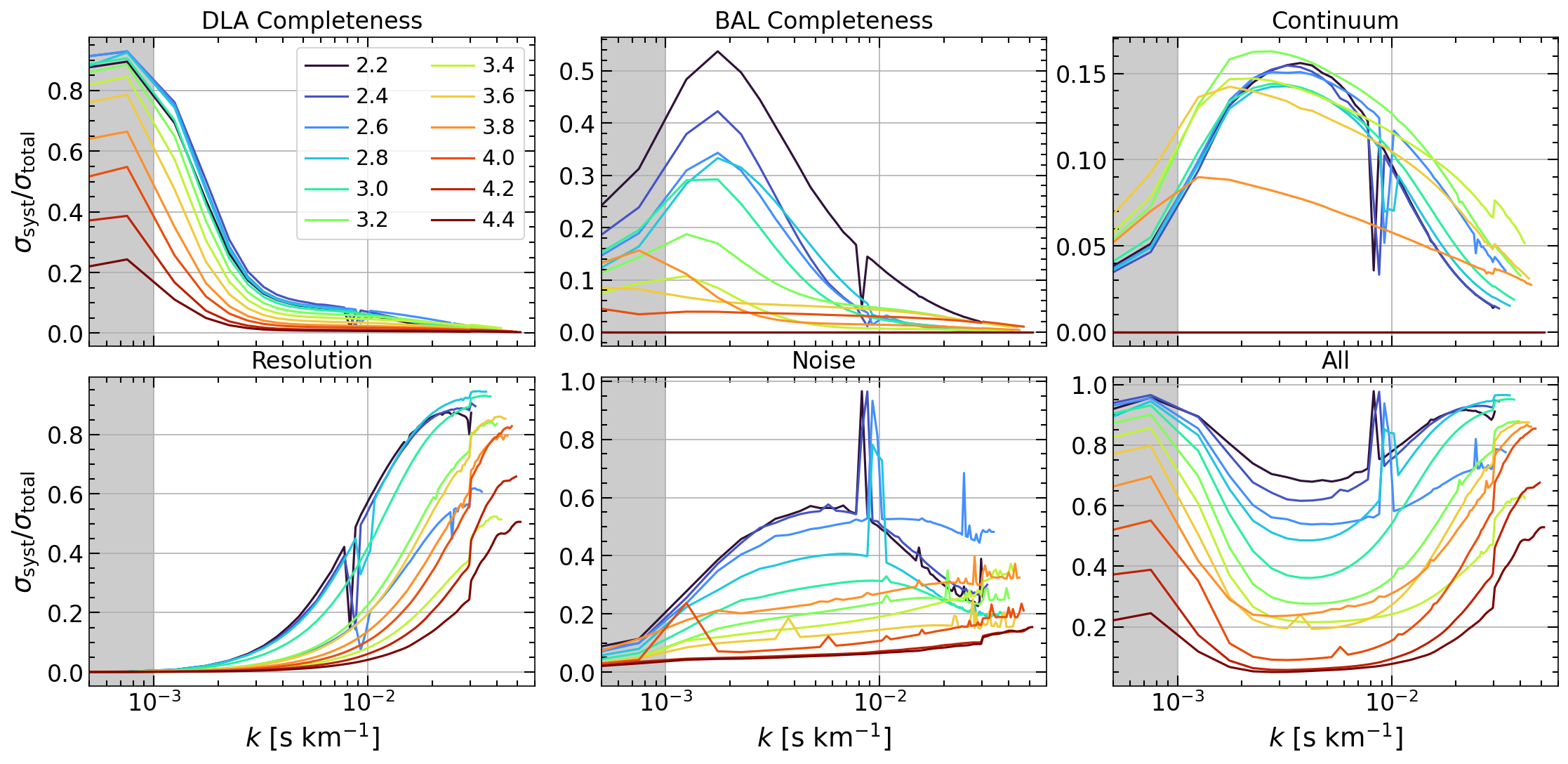}
    \caption{QMLE systematics error budget compared to the total error budget for each redshift bin after correcting for the continuum fitting bias. The noise systematics' \emph{relative} contribution to the total budget decreases due to the increasing impact of resolution systematics. The xQMLE error budget has all of these contributions except for noise systematics.}
    \label{fig:qmle_systematics_breakdown}
\end{figure}

Our systematic error analysis relies on the scalings of the underlying signal. Since the direct application of the estimated power spectrum introduces spurious fluctuations, we instead use a smooth power spectrum $P_\mathrm{smooth}(k, z)$ as described in ref.~\cite{karacayliOptimal1dDesiEdr2023}. Figure~\ref{fig:qmle_systematics_breakdown} presents the overview of our systematic error budget. We present the details below.

\subsubsection{DLA systematics\label{subsec:dla_syst}}
DLA systematics constitute the largest source of systematics on large scales. Their damping wings extend to large wavelength separations, adding power to these scales (low $k$ values). The completeness and purity of the DLA catalog are around 85\%, depending on the confidence level (CL) and $\overline{\mathrm{SNR}}$ cuts \cite{wangDeepLearningDESIDLA2022}. 

\begin{figure}
    \centering
    \includegraphics[width=0.46\linewidth]{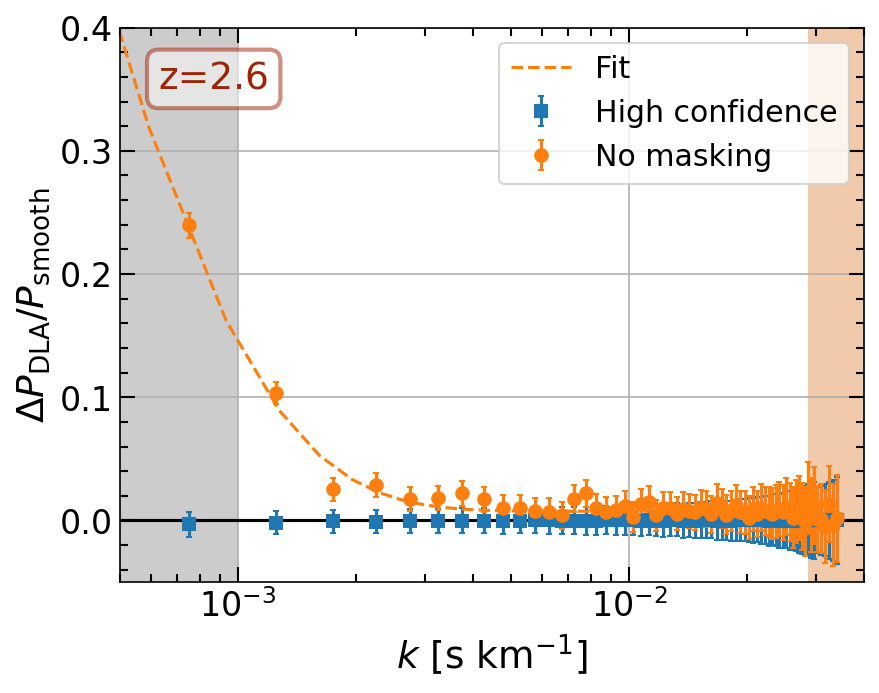}
    \includegraphics[width=0.48\linewidth]{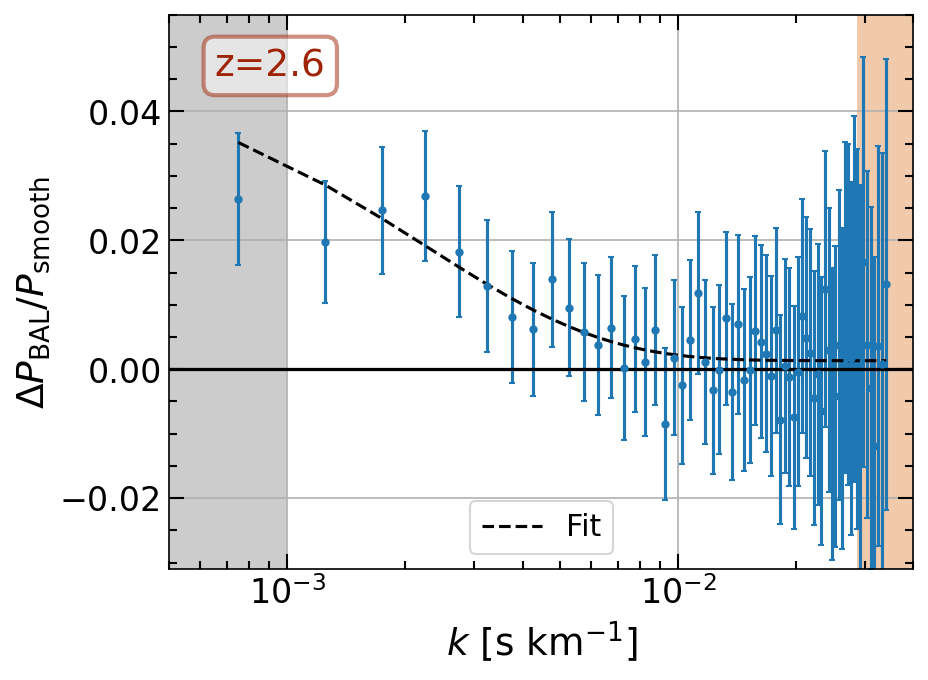}
    \caption{({\it Left}) Relative change in \poned\ when the high-confidence catalog is used ({\it blue squares}) and when DLAs are not masked ({\it orange circles}). The high-confidence results are entirely consistent with our baseline. The systematic error budget is based on a fit to the no-masking case. ({\it Right}) Relative change in \poned\ when BAL features are not masked. These are a factor of ten smaller than the effect of DLAs. The systematic error budget is again based on a fit to this ratio.}
    \label{fig:dla_bal_syst}
\end{figure}

Our baseline DLA catalog is targeted to be more complete; therefore, it is conservative. However, falsely detected DLAs may introduce some bias since false detections can occur in regions with substantial amounts of \lya\ forest absorption. To test this possibility, we rerun the pipeline with a higher confidence catalog where $\overline{\mathrm{SNR}}$ and CL cuts are adjusted to be more granular (see section~\ref{sec:data} and Appendix~\ref{app:dla_cat} for details). Figure~\ref{fig:dla_bal_syst} shows that the results from the high-confidence catalog are entirely consistent with our baseline (conservative strategy) results. This indicates that impurities occur mainly in noisy regions such that they are (1) uncorrelated with the forest and (2) downweighted by the estimator since most of our signal comes from high-SNR spectra (see section~\ref{subsec:highsnr}). As we have mentioned in section~\ref{subsec:data_dla_catalog}, the high-confidence catalog removes a large number of DLAs, but does not have a substantially higher purity. This indicates that it is possible to retain these extra pixels without bias, which is left for future measurements.
The excess power on large scales when DLAs are not masked at $z=2.6$ is also shown in this figure in orange circles.

We estimate our systematics error budget based on scaling these unmasked \poned\ results. A simplified version of the fitting function proposed in ref.~\cite{rogersEffectsOfHcd2018} is adequate to quantify this shape:
\begin{equation}
    \frac{P_\mathrm{1D}^\mathrm{unmasked}}{P_\mathrm{1D}^\mathrm{masked}} = r_\mathrm{DLA}(k, z) = c + \left( \frac{1 + z}{3.0}\right)^\gamma \frac{1}{(a \exp(b k) - 1)^2},\label{eq:dla_func}
\end{equation}
where $\gamma, a, b, c$ are free parameters (they are not redshift dependent as proposed in ref.~\cite{rogersEffectsOfHcd2018}). The best fitting parameters are $\gamma=-0.443, a=1.82, b=690~$\kms$, c=7.49\times10^{-3}$. Based on an estimated 15\% incompleteness ratio in our baseline DLA catalog, our correlated error mode becomes $\sigma^\mathrm{syst}_\mathrm{DLA} = 15\% \, r_\mathrm{DLA}(k, z) P_\mathrm{smooth}(k, z)$.

\subsubsection{BAL systematics}
The metal systems in the quasar outflows are uncorrelated with the diffuse IGM. However, since the absorption lines occur at specific wavelengths, they introduce excess power to the measurement through these ``correlations". QMLE is not complicated by the masking of the associated metal lines compared to the FFT method and can include the high SNR spectra in which these systems are preferentially identified without complications \cite{ennesserMitigationBALquasars2022, martiniDesiBalY12024}. The completeness of the BAL catalog is estimated to be around 80\% for spectra with SNR$>2.5$ \cite{martiniDesiBalY12024}, which constitute by far the majority of the signal, even though they are not the majority of our data set.

Following the same procedure we used to evaluate the DLA systematics, we estimate \poned\ without masking any BAL features. Eq.~\eqref{eq:dla_func} provides a good functional form for these error modes per redshift bin at $k>10^{-3}~$\skm. We fit each redshift bin separately (such that $\gamma$ is omitted) and estimate the correlated BAL error mode with a 20\% incompleteness ratio. The data points and the best-fit curve at $z=2.6$ are shown in figure~\ref{fig:dla_bal_syst}. Note that the BAL contamination ratio is a factor of ten smaller than that for DLAs.

\subsubsection{Continuum fitting systematics\label{subsubsec:continuum}}
Our continuum fitting method biases the quasar continuum estimates towards the average transmission in the forest. This introduces a small bias ($\lesssim 1\%$) that affects all scales, which we failed to detect in our previous work \cite{karacayliOptimal1dDesiEdr2023}. In a companion paper dedicated to validating the \poned\ pipeline \cite{karacayliDesiY1P1dValidation}, we detect and estimate this bias using the stack of 20 synthetic realizations of DR1 up to $z=3.8$. At higher redshifts, this bias is not detectable even with a data set that is 20 times larger than DESI DR1, as shown in the same reference.

The biases $b_\mathrm{cont}(k)$ for each redshift bin are tabulated in ref.~\cite{karacayliDesiY1P1dValidation} and in Appendix~\ref{app:cont_bias}. We apply a correction $P_\mathrm{corr} = P_\mathrm{meas} - \frac{b_\mathrm{cont}(k)}{1 + b_\mathrm{cont}(k)} P_{\mathrm{smooth}}$ for each redshift bin, and assign 30\% of this correction as a correlated error mode. We treat the statistical uncertainties of the estimated biases based on the stack of 20 mocks as an uncorrelated error mode.

\subsubsection{Resolution systematics}
The accurate knowledge of the spectrograph resolution is crucial in extracting the most information from the smallest scales that are valuable in many \lya\ forest \poned\ applications. The DESI spectroscopic pipeline preserves the full native resolution of the 2D spectrograph without degrading the 1D spectrum, yielding an independent resolution matrix for each spectrum. In the companion paper \cite{karacayliDesiY1P1dValidation}, we created 150 tiles of CCD image simulations of just quasars and extracted the 1D spectra of these using the DESI pipeline.  This simulation suite has nearly 675,000 $z>2.3$ quasars, which is 15 times larger than our previous study \cite{karacayliOptimal1dDesiEdr2023}. 
We find that the resolution matrix is valid, although the estimated \poned\ exhibits resolution biases $b_\mathrm{res}$ at an approximately $1.5$\% level. These biases dominate the systematic error at $k\gtrsim 0.01$~\skm.

The scale-dependent error on \poned\ can be written as $b_\mathrm{res} \times 2k^2 R_z^2 \times P_\mathrm{true}(k, z)$, where $R_z \equiv c \Delta\lambda_\mathrm{DESI} / (1 + z) \lambda_\mathrm{Ly\alpha}$. We assign the redshift-dependent bias found in the companion paper as a correlated error mode.
The resolution systematics remain a dominant part of the small-scale error budget, as shown in figure~\ref{fig:qmle_systematics_breakdown}. If a template-based marginalization is desired as done by ref.~\cite{mcdonaldLyUpalphaForest2006}, the theory power spectrum can be multiplied by $e^{2 b_\mathrm{res} k^2R_z^2}$ with $b_\mathrm{res}$ as a free parameter with a Gaussian prior of $\sigma=1.5\%$.

\subsubsection{Noise systematics}
The randomness of observed quasar flux that arises from CCD electron counts adds noise to the measured power spectrum\footnote{We note that this noise power is not related to metal absorption, continuum fitting errors, DLA masking, or noise correlations between pixels.}. The accuracy of this added noise power spectrum is an important source of systematics for DESI, as any error directly propagates to the \poned\ estimates. The DESI pipeline provides estimates of each pixel's variance. However, these pipeline variance estimates suffer calibration errors, which we corrected on average but may differ between CCDs. These variations are the basis of our systematic error budget.

The pipeline variance estimate errors originate at the CCD level. The pipeline divides each CCD into four quadrants and reads each quadrant through separate amplifiers \cite{guySpectroscopicDataProcessingPipeline2022}. The $y$-axis is the wavelength axis, and the transition between amplifiers leaves features in our calibration (as shown in figure~\ref{fig:calib_eta_flux}). The $x$-axis is the fiber direction. Based on the $y$-axis feature, we expect each amplifier to have different noise calibration errors. We consequently performed analyses on data splits based on the amplifier boundaries to quantify the amplifier-to-amplifier uncertainties, which we refer to as amplifier splits.

\begin{figure}
    \centering
    \includegraphics[width=0.48\linewidth]{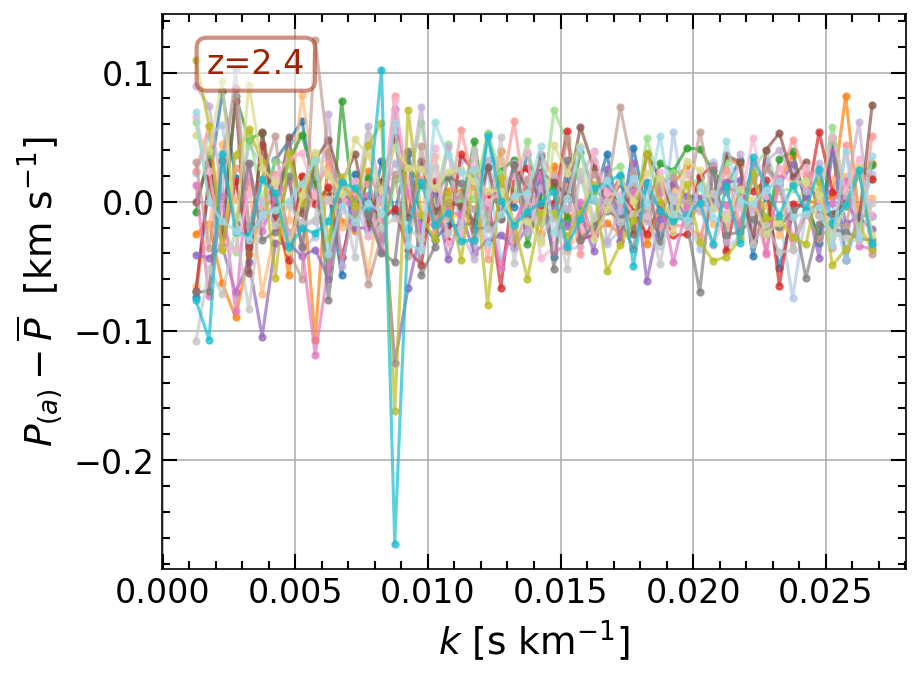}
    \includegraphics[width=0.48\linewidth]{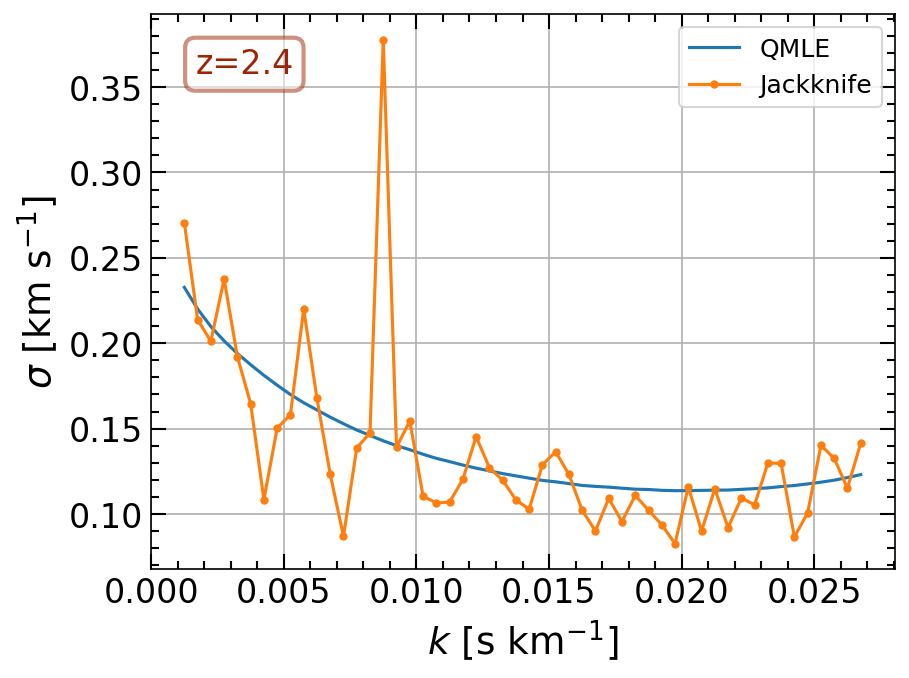}
    \caption{({\it Left}) The difference between one-amplifier-removed Jackknife power spectrum $P_{(a)}$ and the average of all Jackknifes $\overline{P}=\langle P_{(a)} \rangle$ at redshift $z=2.4$. Each color corresponds to a different amplifier. There is a strong feature at $k=0.009$~\skm, which is consistent with a fixed $\mathrm{d}\lambda=9.6$~\AA\ oscillation.
    ({\it Right}) The Jackknife error estimate compared to the statistical errors from QMLE. Having only 20 subsamples makes the former noisy. Our additional systematics error budget is based on the positive difference between these two curves.}
    \label{fig:amplifier_split}
\end{figure}

We divide our data into 20 subsamples based on the amplifier region $a$, of which a spectrum is observed, and estimate 20 one-amplifier-removed Jackknife power spectra $P_{(a)}$. The left panel of figure~\ref{fig:amplifier_split} shows the difference between $P_{(a)}$ and the mean of Jackknife estimates for a single redshift bin.
The strong feature at $k=0.009$~\skm\ shifts in a consistent manner with a fixed $\mathrm{d}\lambda=9.6$~\AA\ ($k_A=0.65$~\AA$^{-1}$) oscillation with respect to redshift. This finding prompted us to examine zero images collected during the timeline of DR1. We find \textit{some} images with quasi-periodic oscillations with $\mathrm{d}\lambda=9.6$~\AA. However, it has quickly become impractical to identify problematic amplifiers when we look through more zero images. First, these features partially cover the amplifier, so the whole amplifier is not defective. Second, defective features are not consistently present across zero images, even on the same night. Third, all amplifiers bear some spurious signal in at least one of the zero images on a given night. For these reasons, we keep the whole data set and increase the systematics error budget based on the Jackknife error estimate of 20 amplifier splits, $\sigma_\mathrm{jack} = \sqrt{\frac{N_a - 1}{N_a} \sum_a (P_{(a)}-\overline{P})^2}$.

The right panel of the same figure compares $\sigma_\mathrm{jack}$ to the statistical errors $\sigma_\mathrm{QMLE}$ at $z=2.4$. The 20 subsamples are insufficient to fully suppress noise in $\sigma_\mathrm{jack}$, which leads to an underestimation of the noise relative to the optimal estimator. Since the optimal estimator gives a Gaussian covariance matrix that is theoretically the smallest, we do not allow decreasing its errors by replacing them with smaller Jackknife estimates.
Instead, we only increase our errors if the Jackknife estimates are larger than the QMLE errors. In other words, we retain only the positive differences in error estimates $\delta \sigma_+$, where $\delta \sigma_+ = \delta \sigma\,\,\mathrm{if}\,\,\delta \sigma = \sigma_\mathrm{jack} - \sigma_\mathrm{QMLE}>0, \mathrm{otherwise}\,\, \delta \sigma_+=0$.

To identify additional outlier features, we compute the mean after removing the two most problematic amplifiers and estimate its error as $\bar{\sigma} = \sigma / \sqrt{17}$, where $\sigma$ is the maximum of Jackknife and QMLE errors. Data points deviating from this mean by more than $\delta P > 3.5 \bar{\sigma}$ are flagged, and their errors are quadratically inflated using $\delta \sigma_+ = (\delta P)^2 / (3.5 \bar{\sigma})$. This procedure identifies one outlier in $z=2.2$, two in $z=2.4$, and three each in $z=2.6\, \text{and}\, 2.8$ bins.

We include $\sigma^\mathrm{syst}_\mathrm{noise, uncorr} = \delta \sigma_+$ as an uncorrelated error mode in the systematics error budget; in other words, $\delta \sigma_+^2$ is added to the diagonals of the covariance matrix. Additionally, based on the scatter of our $\eta$ correction parameter between amplifiers after calibration, we introduce a correlated systematics error mode with $\sigma_\eta =1.5\%$ for $z < 3.7$ and $\sigma_\eta =4\%$ for $z > 3.7$, such that $\sigma^\mathrm{syst}_\mathrm{noise, corr}=\sigma_\eta \times P_N(k, z)$.

Let us briefly explain a failed avenue in improving noise calibration errors. One of our first attempts was to correct each amplifier by its own $\eta_a(\lambda)$ parameter. This correction, calculated in respective SBs, in fact, varies greatly between amplifiers. However, the resulting one-removed Jackknife estimates $P_{(a)}$ become strongly inconsistent when these corrections are applied per amplifier basis. This is because each $\eta_a$ estimate captures amplifier-specific defects that are not necessarily a simple scaling of the noise variance. For example, additive errors or errors coming from the CCD readout cannot be corrected by scaling the combined variance. Averaging over all amplifiers mitigates these amplifier-specific flaws and is more suitable as a scaling correction.

\subsection{Additional data and analysis variations}
In this section, we first examine the 1D correlation function to identify metal transitions that are correlated with \lya\ systems or with other metal systems. These correlations show up as peaks in the correlation function, making them easily identifiable. We then perform and provide an important variation using a high-SNR forest sample ($\mathrm{SNR} > 3$), which is the top 20\% of the entire sample. This variation effectively eliminates the noise power spectrum from the systematics error budget, which could prove advantageous in certain models. We finally perform three sanity check variations (these are not provided): using a narrower rest-frame wavelength range, measuring \poned\ in larger $k$ bins, and smoothing the resolution matrix in the direction parallel to the main diagonal.

\subsubsection{1D correlation functions}
\begin{figure}
    \centering
    \includegraphics[width=0.85\linewidth]{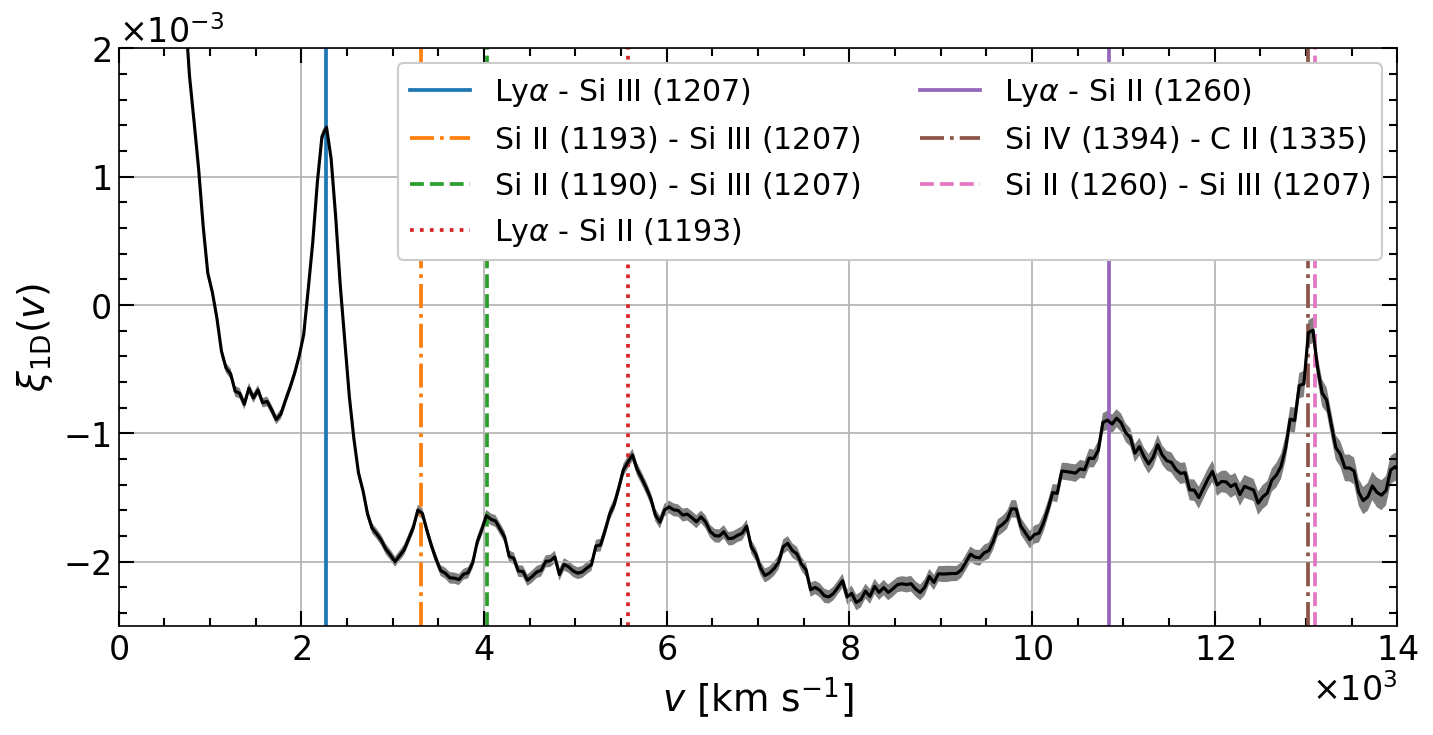}
    \caption{The weighted average of 1D correlation function $\xi_\mathrm{1D}$ across redshift bins of the \lya\ forest. Major metal transitions are highlighted with lines in alternating styles and colors. A complete list of possible transitions is provided in Appendix~\ref{app:xi1d} and shown in figure~\ref{fig:xi1d_lya_all} for the \lya\ forest and figure~\ref{fig:xi1d_sb1_all} for SB1.}
    \label{fig:xi1d_lya_major}
\end{figure}

A strong signature of metal transitions in \lya\ forest measurements is the well-known oscillations in \poned\ and the peaks in the 1D correlation function, $\xi_\mathrm{1D}(v)$. These arise from the fixed velocity separation between two transition lines, which can originate from the doublet transitions of a single metal species, such as C~\textsc{iv} and Si~\textsc{iv}, or from transitions of different species, such as \lya\ and Si~\textsc{iii} (1207~\AA). Figure~\ref{fig:xi1d_lya_major} presents the weighted average $\xi_\mathrm{1D}$ across redshifts of the \lya\ forest ($z_\mathrm{eff}\approx2.4$) and 
highlights the major transitions we identified based on all the combinations of reported transitions in ref.~\cite{pieriCompositeSpectrum2010}. A complete list of possible transitions is provided in Appendix~\ref{app:xi1d} and shown in figure~\ref{fig:xi1d_lya_all} for the \lya\ forest and figure~\ref{fig:xi1d_sb1_all} for SB1.

The well-known \lya-Si~\textsc{iii} (1207~\AA) pair appears as an indubitable peak at $v=2270~$\kms. The weaker peak at $v\approx5500~$\kms\ is due to \lya-Si~\textsc{ii} (1193~\AA), which has historically been difficult to detect in \poned\ analyses \cite{chabanierOnedimensionalPowerSpectrum2019}. More interestingly, there are minor peaks due to several Si~\textsc{ii}-Si~\textsc{iii} correlations. We note that this is an average of all redshift bins, so the signal is significantly boosted. Additionally, while these features are localized in configuration space, they are dispersed in Fourier space. Although accounted for in $\xi_\mathrm{3D}$ analyses \cite{desiKp6BaoLya2024, desiY3LyaBAO2025}, detecting them in individual redshift bins of \poned\ remains difficult. Nevertheless, marginalizing over these features when performing parameter inference would lead to more robust constraints.

Lastly, there are two more prominent peaks at large velocity separations: one at $v=1.1\times 10^4~$\kms\ that arises from \lya-Si~\textsc{ii} (1260~\AA), and another at $v=1.3\times 10^4~$\kms, which could be attributed to Si~\textsc{ii} (1260~\AA)-Si~\textsc{iii} (1207~\AA) and/or Si~\textsc{iv} (1394~\AA)-C~\textsc{ii} (1335~\AA). We observe the latter pair in SB1. The scales of these peaks are comparable to the maximum segment length in our analysis, where each forest segment extends up to $1.8\times 10^4~$\kms. This (1) limits the incidence of two transition lines at such large separations within a single segment and (2) causes the oscillations in \poned\ to be significantly dispersed as their corresponding wavenumber is close to the fundamental frequency. 
As a result, these oscillations are not expected to pose a significant issue for \poned\ analyses. However, they may provide valuable insights into the correlations between these different ionization states and their environment.

While these features offer a promising avenue for studying cosmic metal enrichment \cite{karacayliFrameworkMetals2023}, they are contaminants in \lya\ forest cosmology. These additional correlations between metal species and the \lya\ forest can lead to better models of metal contamination in the forest in future studies.

\subsubsection{High SNR data only\label{subsec:highsnr}}
We performed a high SNR data variation by limiting the sample to $\mathrm{SNR} > 3$ per pixel in the forest region. Here, we do not repeat the entire pipeline, only the \poned\ estimation. This selection reduces the sample to 62,807 quasars and increases the statistical errors by an average of 25\% across redshift and $k$ bins, confirming that most of the signal comes from these high-SNR quasars. Some problematic features are still present due to correlated read noise (e.g., the spike at $k\approx 0.01$~\skm) found in the amplifier splits;  however, we find that the results agree with the baseline results within the statistical uncertainties. The $\chi^2$ between the $\mathrm{SNR} > 3$ sample and the baseline using the bootstrap covariance of the $\mathrm{SNR} > 3$ sample results in $438$ for $680$ degrees of freedom. This test shows we do not have a significant systematic error that depends on the SNR.

We perform an additional variation using this sample by removing BAL quasars identified with the balnicity index while masking the features associated with the absorption index. This removes 3,514 quasars (5.6\% of the $\mathrm{SNR} > 3$ sample) while mitigating the effects of unidentified BAL quasars. The estimated power spectrum effectively remains unchanged, confirming that our masking strategy does not introduce additional errors.

The $\mathrm{SNR} > 3$ sample's advantage is that noise systematics are reduced to insignificant levels. In addition to our baseline measurements, we provide the data points of this high SNR sample. However, the redshift bins $z=4.2$ and $z=4.4$ have 13 and 15 quasars, respectively, which makes the estimated covariance unreliable in these bins. We do not recommend using these redshift bins of the high SNR measurement.

\subsubsection{Shorter rest-frame wavelength range}
Another variation we performed is shortening the quasar rest-frame wavelength range of our analysis to a narrower, more conservative wavelength range of 1070--1160~\AA. This central region of the quasar continuum is farther away from the \lya\ and Ly$\beta$ emission lines of the quasar, providing a smoother region. Additionally, the forest is farther away from the quasar, so this variation is also testing the effect of minor correlations between the forest and its host quasar. We find that this variation also agrees with the baseline results within the statistical uncertainties with $\chi^2=487$ for $754$ degrees of freedom. We conclude that the systematics related to quasar continuum shape and quasar proximity are negligible.

\subsubsection{\texorpdfstring{Larger $k$ binning}{Larger k binning}}
Narrow $k$ bin widths could cause numerical instabilities in the Fisher matrix if not enough Fourier modes enter a bin. We investigate this possibility by increasing the linear $k$ bin size by 50\% ($\Delta k_\mathrm{lin} = 7.5\times 10^{-4}$). This change unavoidably smoothes out fluctuations in \poned, but the outcome remains consistent with baseline and cross-exposure results.


\subsubsection{Smoothing the resolution matrix}
Another numerical instability could occur due to the spectro-perfectionist resolution matrix since it is sensitive to noise in each CCD pixel. Though our results from CCD image simulations do not provide evidence that this effect is important, the real CCDs have hardware defects and cosmic ray contamination that are not included in the simulations. Smoothing every diagonal and off-diagonal element of the resolution matrix in the direction parallel to the main diagonal can mitigate these, so we smooth the resolution matrix with the same window function used in pipeline variance smoothing with a Gaussian window of $\sigma=20$ pixels. This variation again yields consistent results with the baseline.

\section{Discussion\label{sec:discuss}}
DESI DR1 provides an exquisite data set for measuring the \lya\ forest \poned. To harness this data fully, we have conducted a meticulous validation of our measurement pipeline. Using 20 mock realizations of DR1, we identified a $\lesssim 1\%$ bias in \poned\ due to continuum fitting. These biases have been corrected with corresponding systematic error budgets incorporated into our results.
Our analysis demonstrates robustness across variations in SNR, confidence cuts, amplifier regions, and alternative estimators, including the cross-exposure estimator. Importantly, this study precedes the cosmological inference phase, such that none of our analysis choices were influenced by inputs from the parameter inference --- making this a cosmologically blind measurement of \poned.

\subsection{Comparison between methods and previous measurements}
\begin{figure}
    \centering
    \includegraphics[width=\linewidth]{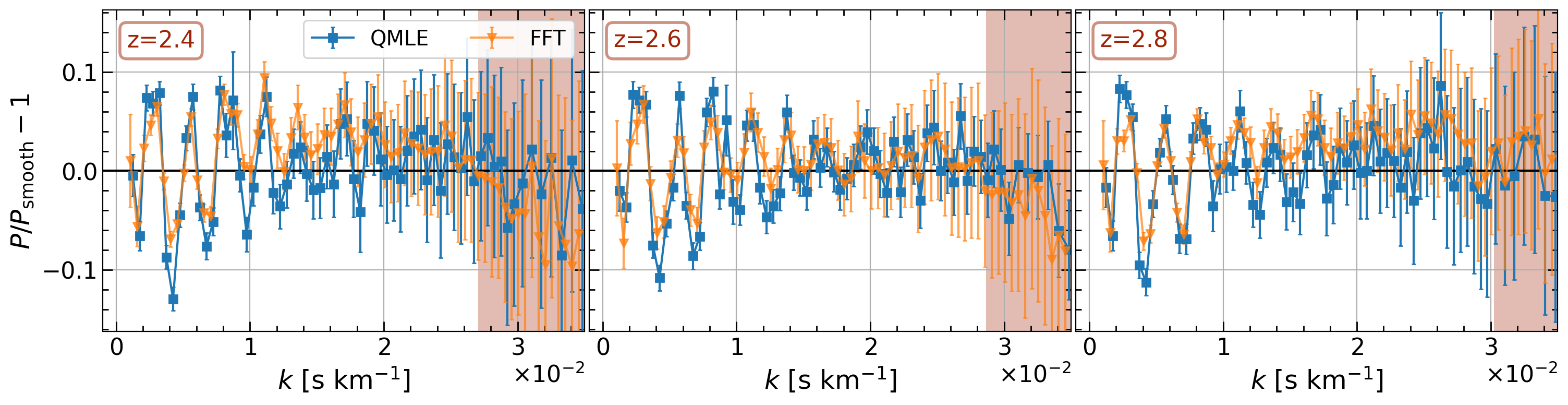} \\
    \includegraphics[width=\linewidth]{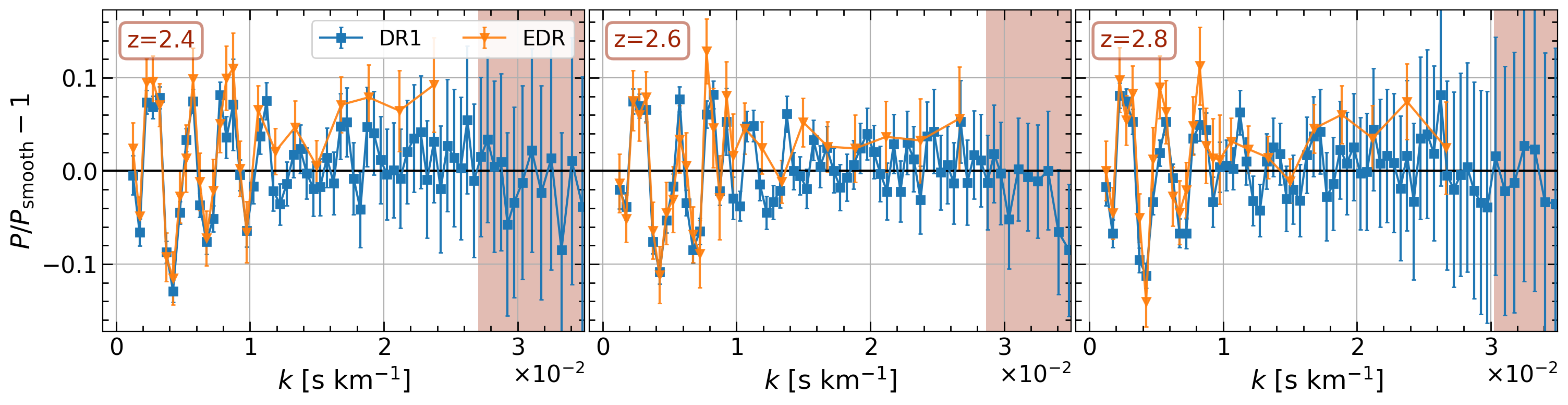}
    \caption{({\it Top}) Estimated \poned\ divided by the smooth power obtained from the QMLE results (interpolated to data points for FFT). The two estimators are in agreement up to half the Nyquist wavenumber (higher $k$ is {\it shaded red}). The differences are due to correlated fluctuations induced by systematics and leftover biases in the FFT due to missing data. Note that the estimators treat the data fundamentally differently, and the samples are not \emph{exactly} the same due to differences in SNR selection and removal of certain quasars with BALs in the FFT sample. ({\it Bottom}) Estimated \poned\ divided by the same smooth power to compare the EDR and DR1 results. The EDR measurement disagrees with DR1 at $k\gtrsim 0.01$~\skm, where EDR has coarser $k$ bins. This coarser $k$ binning complicates a direct comparison. We note that DR1 is processed with improved algorithms and calibration files, such that this disagreement is due to correlated fluctuations induced by noise and resolution systematics, both of which are assigned a smaller error budget in EDR than in DR1. These point to a likely worse performance of the pipeline and an underestimation of systematics in the EDR analysis.}
    \label{fig:compare_qmle_fft_power}
\end{figure}

Figure~\ref{fig:compare_qmle_fft_power} compares our optimal \poned\ measurement to the alternative FFT measurement \cite{ravouxFFTP1dDesiDr12024} in the top panel and shows that both agree up to half the Nyquist wavenumber.
We note that the samples are not \emph{exactly} the same due to differences in SNR selection and removal of certain quasars with BALs in the FFT sample. Besides, the estimators are built with different methods for handling noise, masking, and weighting pixels. As a result, the correlated fluctuations induced by systematics differ between the two methods, making the comparison of \poned\ challenging. The bottom panel compares the EDR and DR1 measurements, which agree at large scales but disagree at $k\gtrsim 0.01$~\skm. Algorithms and calibration files of the spectral processing pipeline have been improved over EDR in DR1 \cite{desiKp2DataRelease12024}. Additionally, a smaller error budget of noise and resolution systematics was estimated in the EDR analysis. These indicate a worse performance of the pipeline and an underestimation of systematics in the EDR analysis. We quantify the level of agreement between measurements below, in section~\ref{subsec:cosmo}, in terms of the forest bias parameter $b_F$.

The difference at large scales in \poned\ between eBOSS and DESI, previously reported in EDR \cite{ravouxFFTP1dEDR2023, karacayliOptimal1dDesiEdr2023}, persists in DR1. The robustness of our measurement is confirmed across various analysis methods and data splits. Additionally, the cross-exposure estimator, which eliminates the pipeline noise contribution and mitigates most instrumental effects on \poned, remains consistent with our baseline measurement, which further reinforces confidence in the DESI pipeline. 

The estimated completeness ratio of 85\% in the DLA catalog \cite{wangDeepLearningDESIDLA2022} is large enough to dominate the error budget at large scales and low redshifts. Also, the shape of DLA contamination in \poned\ aligns with the observed shift between eBOSS and DESI. These suggest that differences in DLA incompleteness and impurity between datasets remain the leading explanation. To investigate this, in the companion FFT measurement paper \cite{ravouxFFTP1dDesiDr12024}, we estimate \poned\ from the common quasars both in eBOSS and DESI while using the DESI DLA catalog for both. The disagreement reduces by a factor of $2-3$, but a $5\%$ offset persists between the two estimated \poned s, indicating errors in spectral extraction are partially responsible for the disagreement. In the companion validation paper, we verified that DESI's spectral extraction does not introduce such biases \cite{karacayliDesiY1P1dValidation}.

Furthermore, we analyzed two DLA catalogs for DESI, one optimized for high completeness and the other for high confidence, and found consistent results between both. This rules out impurities due to confidence cuts in DESI’s DLA catalogs as a potential explanation, but not incompleteness. A notable clue to the disagreement between eBOSS and DESI is the larger disagreement in $z\geq 3.8$ bins, where the reduced mean IGM transmission increases the likelihood of false-positive DLA detections. This suggests that the eBOSS DLA catalog may contain more false positives or more sub-DLAs categorized as DLAs, or that DESI’s catalog is less complete due to the reduced confidence of our DLA finders or more DLAs categorized as sub-DLAs. 

To support this hypothesis, we fit the HCD template model of ref.~\cite{rogersEffectsOfHcd2018} with templates for only small DLAs ($2\times 10^{20}~\text{cm}^{-2} < N_\mathrm{HI} < 10^{21}~\text{cm}^{-2}$) and sub-DLAs ($10^{19}~\text{cm}^{-2} < N_\mathrm{HI} < 2\times 10^{20}~\text{cm}^{-2}$) to the ratio between eBOSS and this work's \poned\ measurements. We avoid complications of ionized silicon oscillations by smoothing both \poned s. We interpolate both to the eBOSS $k$ bin locations and focus on the wavenumber range of $0.001~\text{\skm} <k<0.01~\text{\skm}$.
\begin{figure}
    \centering
    \includegraphics[width=0.8\linewidth]{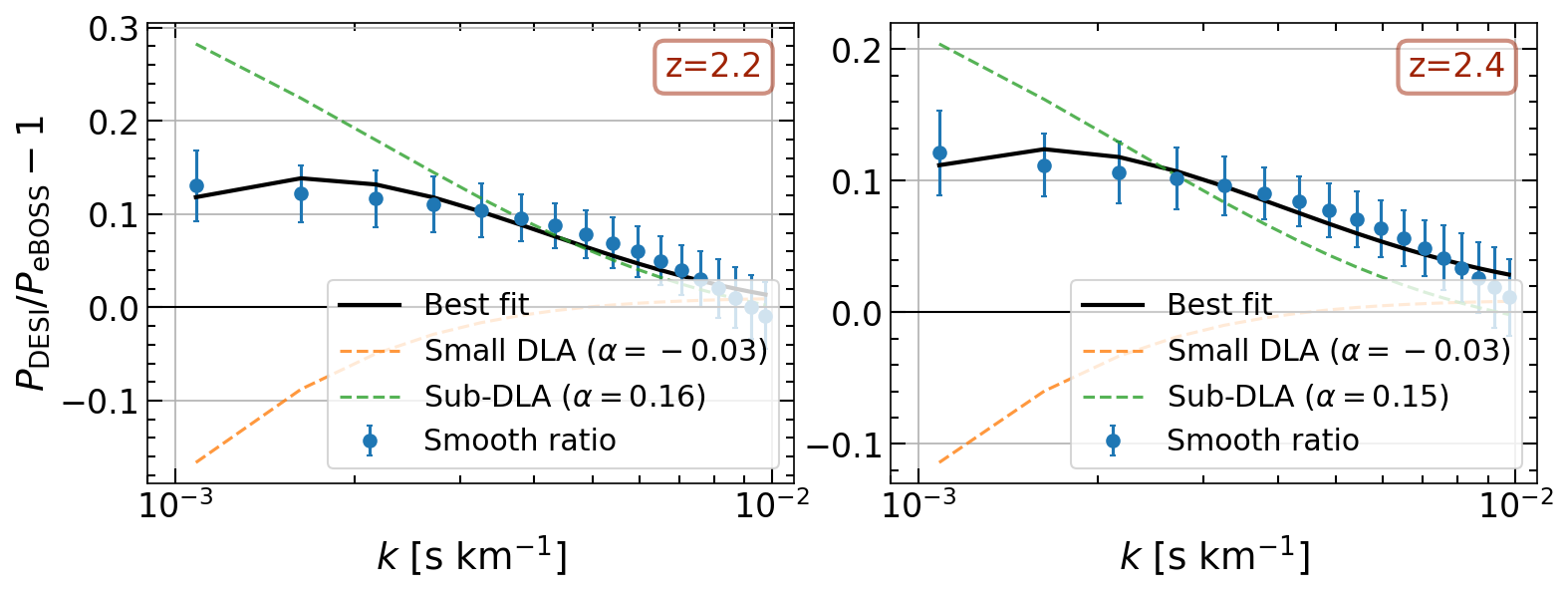}
    \caption{The relative large-scale offset between DR1 optimal \poned\ measurement and eBOSS \poned\ measurement for $ z=2.2\text{ and }2.4$. We fit this quantity with an HCD template model and find that a two-system model can explain the disagreement. This supports our hypothesis that DLA incompleteness and misclassification are the most likely (though not only) explanation.}
    \label{fig:eboss_desi_dla_fit}
\end{figure}
Figure~\ref{fig:eboss_desi_dla_fit} demonstrates the best-fitting HCD template model without any priors on template amplitudes $\alpha$ for $ z=2.2\text{ and }2.4$. A positive $\alpha$ means DESI has more residual objects in the data or fewer identified in the catalog. So the DESI DLA catalog has \emph{fewer} sub-DLAs, or in other words, eBOSS is masking \emph{more} sub-DLAs as DLAs. The large-scale offset is well-captured by this HCD model, but we note that the template amplitudes are overestimated since the offset is partially caused by errors in eBOSS spectral extraction \cite{ravouxFFTP1dDesiDr12024}. However, the HCD model seemingly absorbs these spectral extraction errors. 

Future DESI analyses will keep improving the DLA catalog. DESI DR2 already includes an additional template-based automated search for DLAs, and the optimal combination of three detection methods has been studied in detail \cite{brodzellerConstructionDlaDesiDr2}. We will further improve the DLA incidence and incorporate clustering with the underlying density field in our mocks. These advancements will improve our understanding of the DLA systematics in DESI \poned\ in the future.

\subsection{Comments on the cross-exposure estimator}
The key technical advancement of this work is the development of the optimal cross-exposure estimator formalism. Cross-correlation estimators are widely used in CMB data analyses, primarily because they eliminate the need to model the noise matrix contributing to the covariance in the estimator, which can be arbitrarily correlated. As CCD pixel-level noise properties become increasingly important and remain difficult to model, this estimator holds strong potential for future \poned\ measurements.  

The traditional approach of auto-spectra estimators does not differentiate between the number of exposures for observations with the same SNR. In contrast, xQMLE disregards single observations regardless of their observation time and SNR. Therefore, fully utilizing xQMLE necessitates a different observational strategy---one that maximizes both the number of exposures and the SNR of each exposure. 

That being said, the cross-exposure estimator is not robust to \emph{all} instrumental systematics. One potential systematic error could arise from random errors or systematic biases between spectrographs in wavelength calibration. A velocity shift of $\upsilon_e$ between two exposures could introduce oscillatory features in the estimated power spectrum, yielding $P^*(k) = P(k) \cos(k \upsilon_e)$. If these shifts are randomly distributed between exposures, the oscillations will be partially averaged out according to the probability distribution of $\mathscr{P}(\upsilon_e)$, though some residual oscillations might remain. 

In order for the wavelength calibration errors to be important for \poned, they would have to be larger than 300~\kms\ (4--6~\AA). The DESI wavelength calibration errors are well below this. The largest sources of uncertainty are 
(1) spectrograph misalignment during the night, (2) misalignment between channels (e.g., blue and red), and (3) errors in heliocentric velocity correction. DESI is highly stable, and shifts due to (1) and (2) are estimated to be $\Delta\lambda = 0.025 \text{\AA} \ll 1$~\AA\ \cite[see figure~32 of][]{guySpectroscopicDataProcessingPipeline2022}. For (3), exposures taken of the same quasar in different tiles or different times of the year will have $\ll 1\,$\kms\ errors because the correction to the Solar System barycenter is currently based on the coordinates of the center of the $3^\circ$ diameter field of view. The small errors in this correction are well below the relevant scales. Furthermore, all these effects are sub-Nyquist and cannot actually induce oscillatory features. Therefore, wavelength calibration errors cannot account for the excess fluctuations observed in our xQMLE measurements.

The most likely factor is the limited statistics, as the small number of quasars with sufficient exposures may introduce large fluctuations while leading to an underestimation of statistical errors. A thorough understanding of the cross-exposure estimator requires a detailed study, both through data analysis and dedicated synthetic spectrum pipelines. We defer this investigation to the Year 3 analysis, where the number of $z_\mathrm{qso}>2.1$ quasars with multiple exposures is larger by a factor of 2.5 than in DR1.

\subsection{Measuring the forest bias\label{subsec:cosmo}}
As we mentioned before, the complete cosmological inference is presented in ref.~\cite{chaves-monteroDesiDr1CosmologyP1d} using emulators trained on high-resolution hydrodynamical simulations. In this section, we opt for a simpler approach by using fitting functions. Our formulation results in complete degeneracy between the bias of the \lya\ forest $(b_F)$ and the amplitude of the linear matter power spectrum. Therefore, we fix the cosmological parameters to the best fitting values of the Planck 2018 results \cite{collaborationPlanck2018Results2020} and measure $b_F(z)$.

We employ the fitting function of ref.~\cite{arinyoNonLinearPowerLya2015} for the \lya\ forest 3D power spectrum:
\begin{equation}
    P_\mathrm{3D}(k, \mu) = b_F^2 (1 + \beta_F \mu^2)^2 P_L(k) F_\mathrm{NL}(k, \mu),
\end{equation}
where we find the following reformulation of the non-linear corrections $F_\mathrm{NL}(k, \mu)$ helps the minimizer converge more efficiently:
\begin{equation}
    \ln F_\mathrm{NL}(k, \mu) = q_1 \Delta^2(k) \left[1 - A_\nu^{-1}\frac{(k_\|/k_{\nu, 0})^{b_\nu}}{(k/k_{\nu, 0})^{c_\nu}} \right] - \left( \frac{k}{k_p} \right)^2,
\end{equation}
where $\Delta^2(k) \equiv k^3 P_L(k) / 2\pi^2$, $k_\|\equiv k\mu$, and $k_{\nu, 0} = 1~$\impc. Here, the non-linear growth corrections are quantified by the parameter $q_1$. The pressure smoothing is captured by the parameter $k_p$. The thermal broadening is described with parameters $A_\nu, b_\nu,$ and $c_\nu$.\footnote{We have applied the following transformations to the original fitting function: $c_\nu=b_\nu-a_\nu$ and $A_\nu = k_\nu^{a_\nu}$.} Then, the one-dimensional power spectrum is given by an integral in the transverse direction:
\begin{equation}
    P_\mathrm{1D}(k_\|) = \int~\mathrm{d}\ln k_\bot~\frac{k_\bot^2 P_{\mathrm{3D}}(k_\bot, k_\|)}{2\pi}.
\end{equation}
We integrate this expression from $\ln (10^{-4}~\mathrm{Mpc}^{-1})$ to $\ln (25~\mathrm{Mpc}^{-1})$ using the trapezoidal rule with 500 points.

Additionally, we model the induced oscillations due to major ionized silicon lines and their cross-correlated oscillations as we advocated above, though we find these additional oscillatory features to have minimal impact on fitting. The silicon lines we consider are Si~\textsc{iii}~($1206.51~$\AA), Si~\textsc{ii}~($1190.42~$\AA), Si~\textsc{ii}~($1193.28~$\AA), and Si~\textsc{ii}~($1260.42~$\AA). All Si~\textsc{ii} lines are modeled with the same amplitude parameter $a_{\mathrm{Si~II}}$ for the $1260.42~$\AA\ line, and each transition line is scaled relative to its oscillator strength and wavelength. We describe these in detail in Appendix~\ref{app:metal_model}. Importantly, these oscillations are damped by a Gaussian function as they decorrelate with the \lya\ forest at small scales, such that the total multiplicative correction is $1 + f_\mathrm{metals}(k)\exp(-k^2/2k_s^2)$, where $k$ is in velocity units and $k_s=0.009~$\skm.

If these oscillations are calculated at the exact $k$ value, they can strongly influence the modeling and best-fitting values. In reality, they are smeared out due to the non-zero $k$ bin size. We account for this by averaging the metal corrected \poned\ model in four linearly spaced points for each $k$ bin.

Unfortunately, the \poned\ data alone cannot constrain all the nuisance parameters of the fitting function. For example, the pressure smoothing scale is above what DESI can measure $(k_p\gtrsim10~$\impc$)$ \cite{chabanierAccel2Simulations2024}. Additionally, due to the integration over transverse modes, thermal parameters cannot be untangled without external information. Therefore, we adopt priors derived from state-of-the-art simulations \cite{chabanierAccel2Simulations2024}, referred to as ACCEL2 priors from now on. We fit a polynomial of $\mathcal{Z}=\ln ((1+z)/3.4)$ to the natural logarithm of each parameter of the fitting function and obtain the central values presented in table~\ref{tab:priors}. The two exceptions are $b_F$ and $A_\nu$. The former is the parameter of our interest, so we do not impose any prior on it. For the latter, we find the data strongly rejects the ACCEL2 prior, so we impose no prior on this parameter either. For other parameters, excluding $k_p$, these central values are relaxed with a Gaussian probability based on the reported uncertainties in the same reference. As with any analysis that relies on priors, the measured $b_F$ values might differ if different priors are used.
\begin{table}
    \centering
    \begin{tabular}{|c | c | c|}
    \hline
        Parameter & Central value & Prior \\
        \hline
         $b_\nu$ & 1.64 & $\mathcal{N}(\sigma=0.1)$ \\
         $c_\nu$ &  $\exp(0.2385 + 0.3751 \mathcal{Z}) $ & $\mathcal{N}(\sigma=0.1)$ \\
         $k_p$ & $\exp(3.1398 - 1.5971 \mathcal{Z})$ & Fixed \\
         $\beta_F$ & $\exp(0.4854 - 1.3353 \mathcal{Z}) $ & $\mathcal{N}(\sigma=0.1)$ \\
         $q_1$ & $\exp(-0.2564 + 0.5466 \mathcal{Z} + 5.116 \mathcal{Z}^2)$ & $\mathcal{N}(\sigma=0.1)$\\
         \hline
    \end{tabular}
    \caption{ACCEL2 priors of our analysis. The central values are obtained by fitting a polynomial of $\mathcal{Z}=\ln ((1+z)/3.4)$ to the natural logarithm of each parameter reported in ref.~\cite{chabanierAccel2Simulations2024}. Since the pressure smoothing scale $k_p$ is above what DESI can measure, it is fixed. For other parameters, priors are relaxed with a Gaussian based on the reported uncertainties in the same reference. They are illustrated in figure~\ref{fig:accel2_priors} in Appendix~\ref{app:metal_model}.}
    \label{tab:priors}
\end{table}

\begin{figure}
    \centering
    \includegraphics[width=0.48\linewidth]{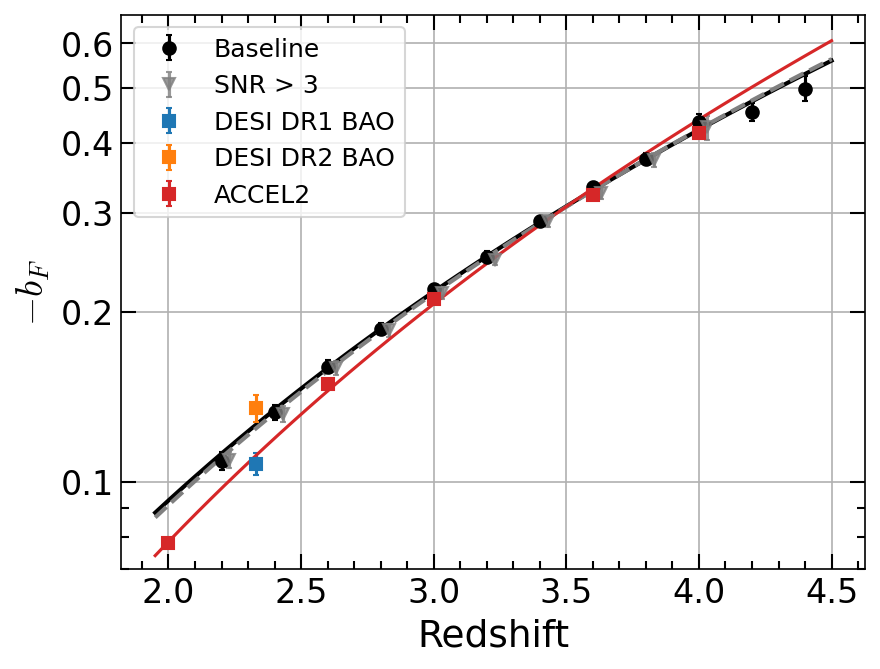}
    \includegraphics[width=0.48\linewidth]{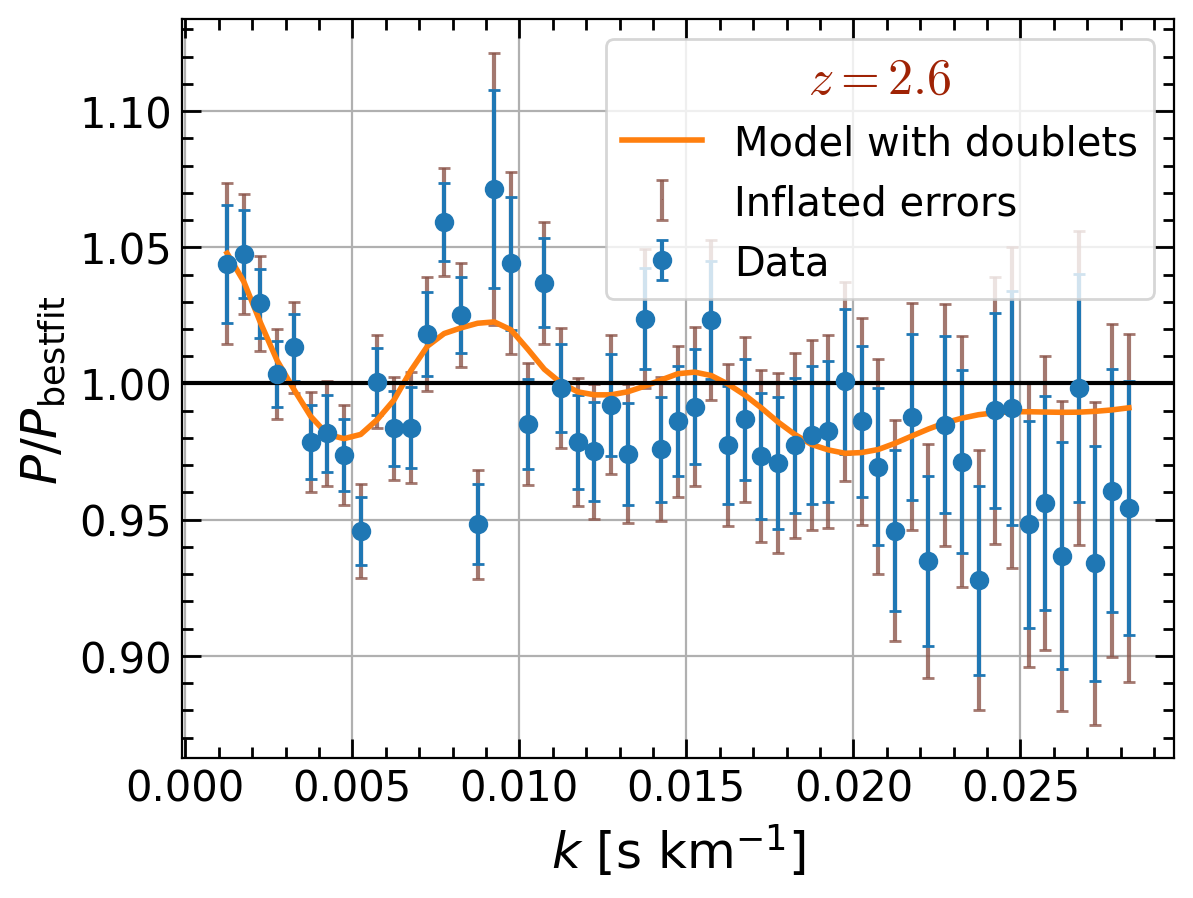}
    \caption{({\it Left}) $b_F$ measurement in each redshift bin. Our findings from the baseline and $\mathrm{SNR}>3$ samples are consistent. They also agree with the DESI DR2 \lya\ BAO measurement while disagreeing with DR1 BAO and ACCEL2 simulation. ({\it Right}) Power spectrum divided by the best-fitting model without doublets and inflated errors $(P_\mathrm{bestfit})$ for $z=2.6$. The best-fitting model with doublets ({\it orange line}) can explain the oscillatory feature at $k<0.01~\text{\skm}$, though outliers drive the reduced chi-squared values $(\chi/\nu)$ to larger than one. We inflate the errors on all data points with an ad-hoc factor to reach $\chi/\nu\approx 1$ instead of cherry picking data points.}
    \label{fig:bias_beta}
\end{figure}

Let us now introduce some hindsight information into our procedure. Our initial best-fitting models had high chi-squared values in almost all redshift bins. The problem seems to stem from a combination of oscillatory features, underestimated errors for certain outliers in measured \poned, and our model's inability to capture small-scale behavior.\footnote{Even though our \lya\ forest model is simplistic and may contribute to the poorness of fit, we expect the major reasons are underestimated errors and inability to model oscillatory features due to unknown metals, calibration errors, and/or other artefacts of the pipeline.} First, we notice that the chi-square value significantly improves (though remains high) if residual metal doublets of Mg~\textsc{ii} and C~\textsc{iv} are marginalized by fitting a free amplitude parameter (we find no evidence for a N~\textsc{v} doublet residual). For this purpose, we adopt the simpler fitting function developed in ref.~\cite{karacayliFrameworkMetals2023} for each doublet $d$ we consider: $P_d = a_d (1.25 + \cos(kv_d))\exp(-k^2 / 2k_s^2)$, where $k_s$ is the same Gaussian dampening scale as above and $v_d$ is the doublet separation in velocity units: $v_\mathrm{Mg~II} = 768.6~\text{\kms}$ and $v_\mathrm{C~IV} =  497.6~\text{\kms}$. Note that this is an additive power, unlike the multiplicative silicon oscillations. The reduction in $\chi^2$ is most dramatic in $z\leq 2.8$, ranging from $\Delta \chi^2 = -24$ at $z=2.8$ to $\Delta \chi^2 = -80$ at $z=2.2$. Second, unfortunately, we have to resort to inflating the diagonals of the covariance matrix to mitigate the bad chi-square value. We use an ad-hoc value of $(\chi^2/\nu + 1) / 2$ to multiply the diagonal of the covariance matrix, which achieves $\chi^2/\nu\sim 1$ with a tendency to overfit, where $\nu$ is the number of degrees of freedom. These changes are illustrated in the right panel of figure~\ref{fig:bias_beta} for $z=2.6$. Notably, the $b_F(z)$ measurement, as well as silicon parameters, remain largely unaffected by these changes.

We measure $b_F$ in each redshift bin independently using the minimizer \texttt{iminuit} \cite{jamesMinuit1975, iminuit}. Note that we incorporate the systematics through correlated fluctuations in the covariance matrix, which is equivalent to fitting a free amplitude parameter for a given template. Figure~\ref{fig:bias_beta} shows our $b_F(z)$ measurements from both baseline and $\mathrm{SNR}>3$ samples in the left panel.
They are consistent with each other. Furthermore, DESI DR2 \lya\ BAO analysis measures the same bias parameter at an effective redshift $z_\mathrm{eff}=2.33$, which agrees with our findings. However, DR1 BAO analysis disagrees with DR2 analysis and our findings. ACCEL2 simulation seems to produce smaller forest biases than our measurement at $z<3.0$, which agrees with DR1 BAO but disagrees with our findings and DR2 BAO. 
$\beta_F$ results from both baseline and high-SNR samples are driven by the prior across all redshifts. The remaining nuisance parameters are discussed in Appendix~\ref{app:metal_model}.

We strongly detect Si~\textsc{iii} oscillations at $z<3.9$ with $0.03<a_\mathrm{Si~III}<0.065$ and Si~\textsc{ii} oscillations at $z<3.5$ with $0.025<a_\mathrm{Si~II}<0.045$. The Si~\textsc{iii} detection gradually vanishes with redshift, whereas the Si~\textsc{ii} detection suddenly drops to zero at $z=3.6$. The quantity $f=a\times[1 - \overline{F}(z)]$ is typically modeled to be a constant with a fiducial choice of mean IGM transmission $\overline{F}(z)$ \cite{chabanierOnedimensionalPowerSpectrum2019, waltherEmulatingLyaForestEboss2024}. Using the $\overline{F}(z)$ measurement of ref.~\cite{turnerLyaForestMeanFluxFromDesiY12024}, we find that both $f_\mathrm{Si~III}$ and $f_\mathrm{Si~II}$ increases with redshift. These parameters partly fit the spurious fluctuations in our data, which deters us from interpreting their redshift evolution with IGM thermodynamics. However, they offer insight for analyses that treat these as nuisance parameters. We recommend relaxing the non-evolving $f$ parameterization in future analyses.

Let us fit a power law to our $b_F(z)$ measurements and use this framework to compare our optimal estimator results to the FFT measurement. An inferred parameter comparison is more suitable since it is challenging to disentangle correlated systematics between \poned\ measurements. The power law proves to be an exceptionally good fit with the following best-fitting values for our baseline analysis:
\begin{equation}
    b_F(z) = b_{F, 0}\left(\frac{1 + z}{4}\right)^\gamma = (-0.2175\pm0.0018)\left(\frac{1 + z}{4}\right)^{2.963\pm0.058},
\end{equation}
where the errors are computed via the Monte Carlo method. Turning our focus back to comparing the FFT measurement \cite{ravouxFFTP1dDesiDr12024}, we apply the same inference analysis to it.
We find that the FFT measurement prefers smaller $b_F$ across all redshifts with the best-fitting values of $b_{F, 0} = -0.2135\pm0.0018$ and $\gamma = 2.955\pm0.060$. The redshift evolution of the bias is in perfect agreement between estimators, while the $b_{F, 0}$ parameter is approximately 2.2 sigma away (assuming both estimates are maximally correlated, which is not exactly true due to differences between methods). This potentially indicates that the measurements of the linear matter power spectrum amplitude might disagree at this level (ignoring modeling errors).

The oscillatory parameters differ the most between FFT and optimal measurements. Si~\textsc{iii} oscillations are detected with a lower value of $a$ in the FFT measurement than our optimal measurement, whereas Si~\textsc{ii} remains undetected with zero amplitude. This is likely due to the uncorrected smearing caused by the survey window function in the FFT results. However, both FFT and optimal measurements prefer comparable amplitudes of Mg~\textsc{ii} and C~\textsc{iv} doublet residuals. We recommend accounting for these when using both \poned\ measurements.

This 2.2 sigma $b_{F, 0}$ disagreement has a low 3\% probability, which is indicative of a systematic difference between methods. Our model is simplistic, so it could partially explain the disagreement. There could be leftover systematics in the FFT measurement due to missing data. Since the smearing of oscillations and other correlated systematics makes the direct comparison of \poned\ between FFT and this work difficult, this disagreement will be investigated in future data releases and/or in a future work that presents a full cosmological analysis.

Applying the same pipeline to EDR results, we find that it prefers a larger $b_F$ across all redshifts with the best-fitting values of $b_{F, 0} = -0.2239\pm0.0024$ and $\gamma = 3.024\pm0.090$. This is 2.1 sigma larger than the DR1 value, assuming the data sets are uncorrelated. As we identified in the previous section, this is likely sourced by the worse performance of the pipeline and underestimated and undercontrolled noise and resolution systematics in the EDR analysis. Limiting the $b_F$ fitting to $k<0.01~$\skm\ range eliminates the discrepancy though it also weakens the constraints.

Lastly, the degeneracy between $b_F$ and the amplitude of the linear matter power spectrum can be broken with additional statistics such as cross-correlations between \poned\ and the lensing maps from the cosmic microwave background \cite{karacayliCmbLensingLyaP1d2024}. This requires a theoretical model that can jointly describe both statistics, such as the effective field theory of large-scale structure \cite{ivanovEffectiveLya2024, ivanovLyaForestEfteBOSS2025}.

\subsection{Recommendations for use}
The following steps should be taken when using our measurement to constrain models:
\begin{enumerate}
    \item Apply the redshift-dependent $k$ scale cut: $10^{-3}~$\skm$< k < 0.5 \pi / R_z$, where $R_z \equiv c \Delta\lambda_\mathrm{DESI} / (1 + z) \lambda_\mathrm{Ly\alpha}$ and $\Delta\lambda_\mathrm{DESI}=0.8$~\AA.
    \item Do not use redshift bins of $4.2$ and $4.4$ of the high-SNR sample since the covariance matrix is not reliable due to low statistics in these bins.
    \item Use the full covariance matrix. The correlated template marginalization handles the systematics we discussed without extra parameters. Additional templates for small DLAs and sub-DLAs can be introduced for conservative analyses \cite{rogersEffectsOfHcd2018}. If an alternative marginalization is desired for the resolution systematics, remove its contribution from the covariance matrix and multiply the theory power spectrum by $e^{2b_\mathrm{res}k^2R_z^2}$ and apply a Gaussian prior of $\sigma=1.5\%$ to $b_\mathrm{res}$.
    \item Model both Si~\textsc{iii} and Si~\textsc{ii} oscillations as well as Mg~\textsc{ii} and C~\textsc{iv} residual doublet contaminations. Analyses considering all redshift bins using a single $f$ parameter (instead of an independent $a$ parameter for each redshift bin) for ionized silicon oscillations are likely to benefit from relaxing the constant $f$ assumption.
    \item If a bad $\chi^2_\nu$ is encountered, which is likely to happen, inflate the diagonals of the covariance matrix until you obtain $\chi^2_\nu\sim 1$. Our ad-hoc additive diagonal term in the forest bias measurement is redshift-dependent, which we provide separately, and can be included from the beginning if desired.
\end{enumerate}

\section{Summary\label{sec:summary}}
The 1D \lya\ forest \poned\ measures the clustering of neutral hydrogen in the IGM and provides constraints on the sum of neutrino masses, warm dark matter models, and the thermal state of the IGM. The DESI DR1 \lya\ quasar sample has over 300,000 spectra suitable for \poned\ analysis, surpassing its predecessor, eBOSS, by a factor of 1.7. DESI also has higher resolution and improved calibration and noise estimation in its spectroscopic pipeline.
To fully exploit the corresponding increase in statistical power, we employed the optimal estimator to measure \poned\ and conducted a rigorous investigation of systematics. A major advancement of our work is extending the optimal estimator formalism to a cross-exposure estimator, which removes the need to model the noise matrix.

The optimal estimator is robust against gaps in spectra, such that the space of systematics is limited to five sources: DLA incompleteness, BAL incompleteness, continuum fitting errors, spectrograph resolution errors, and miscalibrated and correlated noise. We applied corrections only for continuum fitting biases, which are clearly detected in simulated spectra. We conservatively include 30\% of these corrections in the systematic error budget.

We modeled the systematic errors as correlated fluctuations in most cases. In other words, we incorporated shape information into the covariance matrix to isolate the corresponding error modes. We found that the systematics dominate relative to the statistical errors at low redshifts, $z \leq 2.6$. For DESI’s medium-SNR spectra, the noise power spectrum is an important source of systematics. Beyond percent-level miscalibrations, we identified troubling quasi-periodic features arising from CCD read noise using amplifier splits. We identified the strongest features and inflated the errors on the most affected modes. These are the only uncorrelated fluctuations in our covariance matrix. We provide a decomposition of our systematic error budget to enable alternative treatments in cosmological inference.

Our \poned\ measurement is in good agreement with DESI EDR results at large scales. As a result, the previously reported disagreement between eBOSS and DESI measurements persists in DR1. We confirmed the robustness of our measurement through analysis variations and data splits, including a high-confidence variation on our DLA catalog. In the companion FFT paper, we showed that some of this disagreement is partially attributed to the spectral extraction errors. In this work, we showed that a template model for the sub-DLA contamination can fit the disagreement at large scales, seemingly also fitting for the spectral extraction errors. The software packages, including the continuum fitting and DLA finder packages, have evolved since the eBOSS measurement. Reanalyzing eBOSS spectra with these updated tools will be fruitful in understanding the discrepancy in the future.

In conclusion, this work is the most precise measurement of \poned\ to date and marks a significant advancement in the technical tools addressing the challenges of the \lya\ forest now and in future analyses. The collective DESI \poned\ analysis includes rigorous validation tests \cite{karacayliDesiY1P1dValidation} and an alternate measurement using FFTs \cite{ravouxFFTP1dDesiDr12024}. The cosmological inference will be presented in the following papers \cite{chaves-monteroDesiDr1CosmologyP1d}.

\paragraph{Data Availability.}
All data points shown in the figures are available on the following website: \url{https://doi.org/10.5281/zenodo.16943723}. DR1 is accessible to the public at \url{https://data.desi.lbl.gov/doc/releases/dr1/}.

\paragraph{Software.} Our estimator\footnote{\url{https://github.com/p-slash/lyspeq}} is written in \texttt{c++}.
It depends on \texttt{CBLAS} and \texttt{LAPACKE} routines for matrix/vector operations, \texttt{GSL}\footnote{\url{https://www.gnu.org/software/gsl}} for certain scientific calculations \citep{GSL}, \texttt{FFTW3}\footnote{\url{https://fftw.org}} for fast Fourier transforms \citep{FFTW05}; and uses the Message Passing Interface (MPI) standard\footnote{\url{https://www.mpich.org}} to parallelize tasks.
The quasar spectra are organized with the \texttt{HEALPix} \citep{healpix} scheme on the sky.
We use the following commonly-used software in \texttt{python} analysis: \texttt{astropy}\footnote{\url{https://www.astropy.org}}
a community-developed core \texttt{python} package for Astronomy \citep{astropy:2013, astropy:2018, astropy:2022},
\texttt{numpy}\footnote{\url{https://numpy.org}}
an open source project aiming to enable numerical computing with \texttt{python} \citep{numpy},
\texttt{scipy}\footnote{\url{https://scipy.org}} an open-source project with algorithms for scientific computing.
\texttt{healpy} to interface with \texttt{HEALPix} in \texttt{python} \citep{healpy},
Finally, we make plots using
\texttt{matplotlib}\footnote{\url{https://matplotlib.org}}
a comprehensive library for creating static, animated, and interactive visualizations in \texttt{python}
\citep{matplotlib}.

\acknowledgments
This material is based upon work supported by the U.S. Department of Energy (DOE), Office of Science, Office of High-Energy Physics, under Contract No. DE–AC02–05CH11231, and by the National Energy Research Scientific Computing Center, a DOE Office of Science User Facility under the same contract. Additional support for DESI was provided by the U.S. National Science Foundation (NSF), Division of Astronomical Sciences under Contract No. AST-0950945 to the NSF’s National Optical-Infrared Astronomy Research Laboratory; the Science and Technology Facilities Council of the United Kingdom; the Gordon and Betty Moore Foundation; the Heising-Simons Foundation; the French Alternative Energies and Atomic Energy Commission (CEA); the National Council of Humanities, Science and Technology of Mexico (CONACYT); the Ministry of Science and Innovation of Spain (MICINN), and by the DESI Member Institutions: \url{https://www.desi.lbl.gov/collaborating-institutions}.

The DESI Legacy Imaging Surveys consist of three individual and complementary projects: the Dark Energy Camera Legacy Survey (DECaLS), the Beijing-Arizona Sky Survey (BASS), and the Mayall z-band Legacy Survey (MzLS). DECaLS, BASS and MzLS together include data obtained, respectively, at the Blanco telescope, Cerro Tololo Inter-American Observatory, NSF’s NOIRLab; the Bok telescope, Steward Observatory, University of Arizona; and the Mayall telescope, Kitt Peak National Observatory, NOIRLab. NOIRLab is operated by the Association of Universities for Research in Astronomy (AURA) under a cooperative agreement with the National Science Foundation. Pipeline processing and analyses of the data were supported by NOIRLab and the Lawrence Berkeley National Laboratory. Legacy Surveys also uses data products from the Near-Earth Object Wide-field Infrared Survey Explorer (NEOWISE), a project of the Jet Propulsion Laboratory/California Institute of Technology, funded by the National Aeronautics and Space Administration. Legacy Surveys was supported by: the Director, Office of Science, Office of High Energy Physics of the U.S. Department of Energy; the National Energy Research Scientific Computing Center, a DOE Office of Science User Facility; the U.S. National Science Foundation, Division of Astronomical Sciences; the National Astronomical Observatories of China, the Chinese Academy of Sciences and the Chinese National Natural Science Foundation. LBNL is managed by the Regents of the University of California under contract to the U.S. Department of Energy. The complete acknowledgments can be found at \url{https://www.legacysurvey.org/}.

Any opinions, findings, and conclusions or recommendations expressed in this material are those of the author(s) and do not necessarily reflect the views of the U. S. National Science Foundation, the U. S. Department of Energy, or any of the listed funding agencies.

The authors are honored to be permitted to conduct scientific research on Iolkam Du’ag (Kitt Peak), a mountain with particular significance to the Tohono O’odham Nation.

\appendix
\section{Details on the DLA catalog\label{app:dla_cat}}
\begin{figure}
    \centering
    \includegraphics[width=0.9\linewidth]{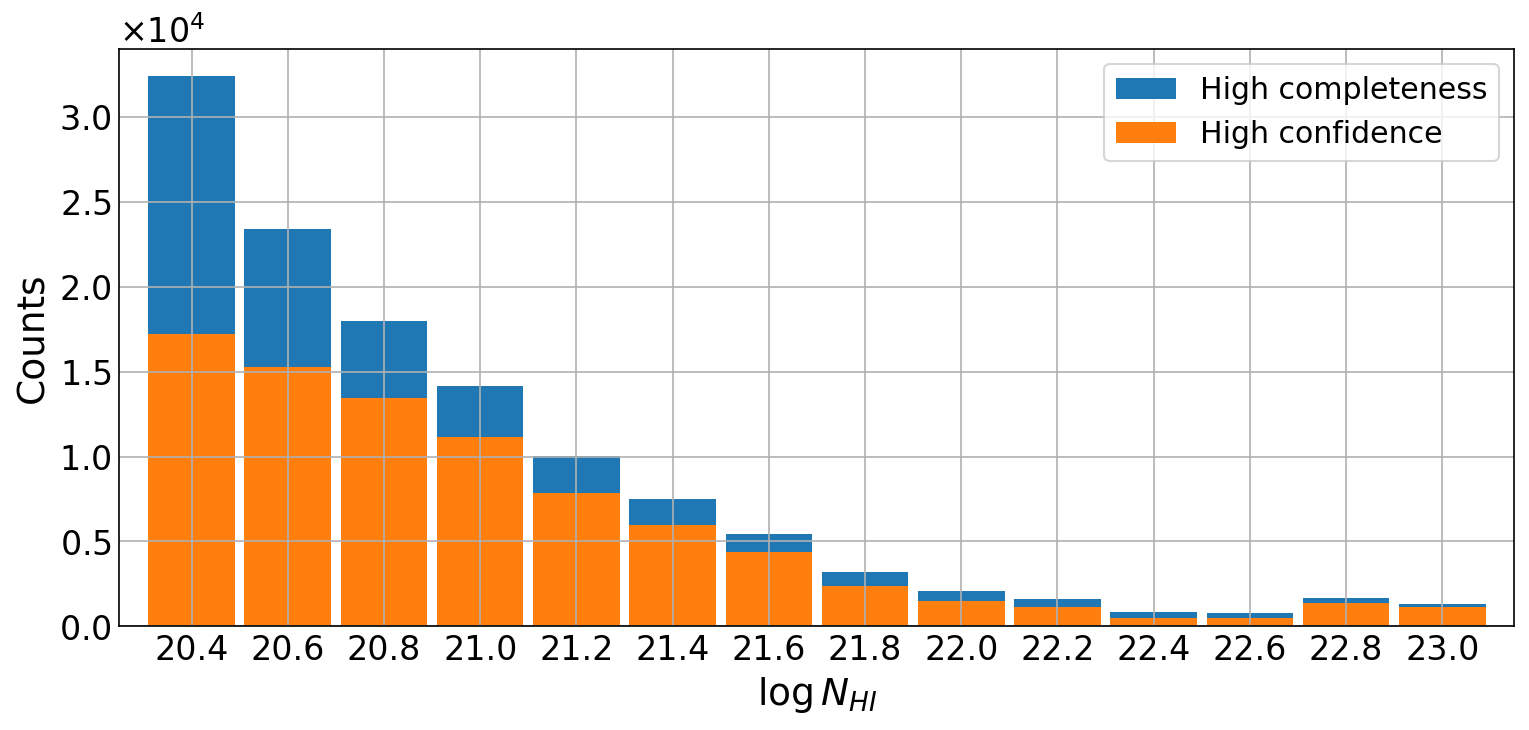}
    \caption{Number of systems in each DLA catalog per column density estimated by the DLA finders in the concordance catalog.}
    \label{fig:dla_nhi_compare_catalogs}
\end{figure}

We select DLA systems based on the average signal-to-noise ratio $\overline{\mathrm{SNR}}$ per pixel between 1420--1480~\AA\, in the quasar rest frame. The high-completeness catalog follows the confidence cuts in EDR analysis \cite{karacayliOptimal1dDesiEdr2023, ravouxFFTP1dEDR2023}: All systems that the GP DLA finder identifies are kept regardless of $\overline{\mathrm{SNR}}$. The systems that CNN identifies are retained if the host quasar spectrum has $\overline{\mathrm{SNR}}>3$ or if the confidence level (CL) is greater than 0.3. This high-completeness catalog has 122,458 DLAs in 85,186 quasars, resulting in an average of 1.4 DLAs per sightline with at least one DLA.
The high-confidence catalog applies CL$>0.9$ for GP systems and CNN CL cuts tiered at multiple $\overline{\mathrm{SNR}}$ values based on ref.~\cite{wangDeepLearningDESIDLA2022}: $0 < \overline{\mathrm{SNR}} \leq 1$ requires CL$> 0.5$, $1 < \overline{\mathrm{SNR}} \leq 2$ requires CL$> 0.4$, $2 < \overline{\mathrm{SNR}} \leq 3$ requires CL$> 0.3$, and $\overline{\mathrm{SNR}}>3$ requires CL$> 0.2$. This high-confidence catalog has 83,795 DLAs in 57,909 quasars, which removes 38,663 DLAs from the high-completeness catalog but has the same average of 1.4 DLAs per sightline. Figure~\ref{fig:dla_nhi_compare_catalogs} compares the number of systems as a function of column density.

\section{Continuum fitting bias corrections\label{app:cont_bias}}
Besides contaminating low $k$ modes, the continuum fitting adds extra power to all scales through three-point correlations even when the polynomials of $\ln\lambda$ are marginalized out in the optimal estimator method.
In ref.~\cite{karacayliDesiY1P1dValidation}, we calculate these biases due to continuum fitting errors using the stack of 20 synthetic realizations of DR1. We tabulate these bias values here for reference in table~\ref{tab:bias_corrections}.
\begin{table}
    \centering
    \begin{tabular}{|c|c || c | c || c | c|}
        \hline
        $z$ & $b_\mathrm{cont}(k) \times 10^3$ & $z$ & $b_\mathrm{cont}(k) \times 10^3$ & $z$ & $b_\mathrm{cont}(k) \times 10^3$ \\
        \hline
        2.2 & $6.768 - 0.699 x$ & 2.8 & $5.611 - 0.098 x$ & 3.4 & $8.913 + 0.093 x$ \\
        2.4 & $5.736 - 0.692 x$ & 3.0 & $6.206 - 0.101 x$ & 3.6 & $10.825 + 0.172 x$ \\
        2.6 & $5.865 - 0.026 x$ & 3.2 & $8.335 - 0.089 x$ & 3.8 & $7.599 - 0.059 x$ \\
        \hline
    \end{tabular}
    \caption{Bias corrections $b_\mathrm{cont}(k)$ that are derived from the stack of 20 fully contaminated mocks, where $x \equiv k/k_0$ and $k_0=0.009~$\skm.}
    \label{tab:bias_corrections}
\end{table}

\section{1D Correlation functions\label{app:xi1d}}
\begin{figure}
    \centering
    \includegraphics[width=0.9\linewidth]{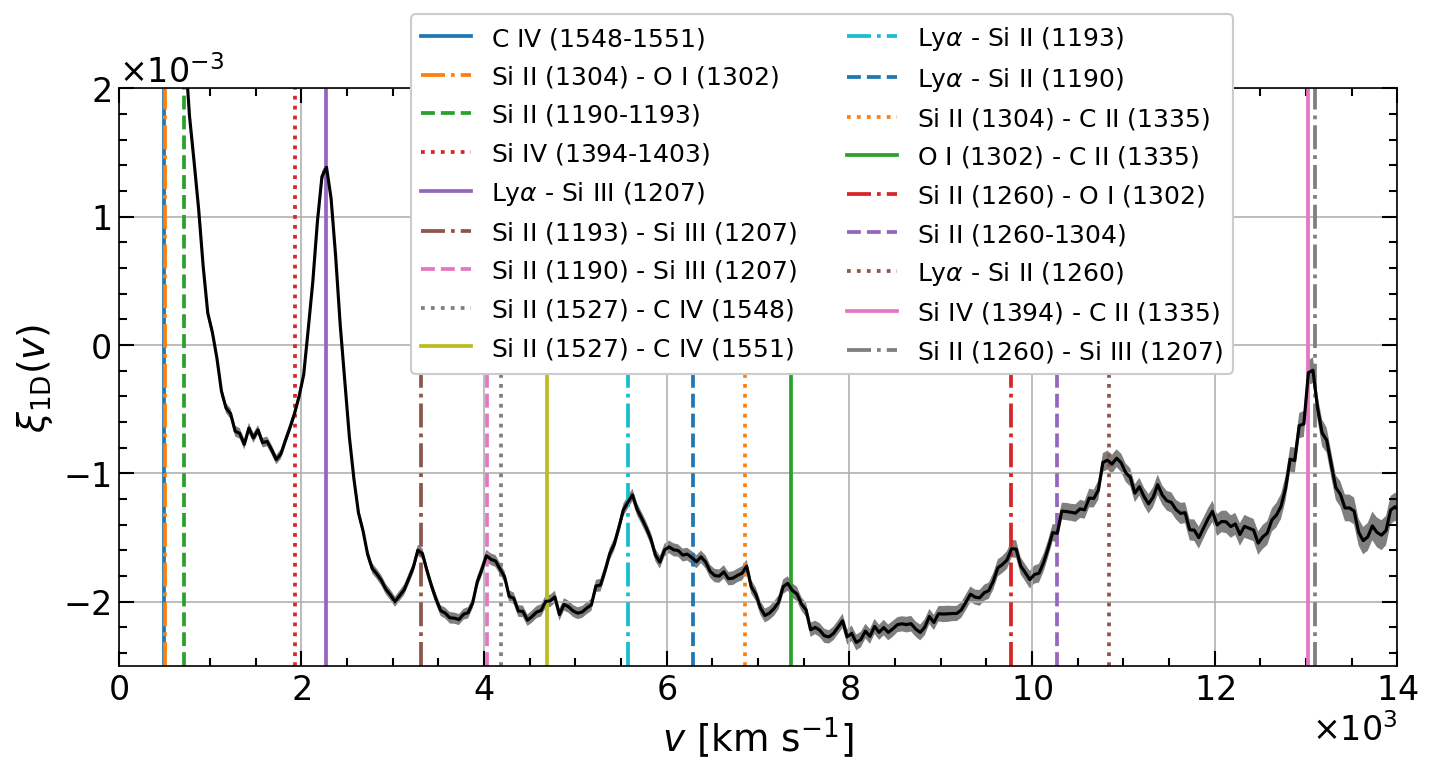}
    \caption{The weighted average of the 1D correlation function $\xi_\mathrm{1D}$ across redshift bins of the \lya\ forest (same as figure~\ref{fig:xi1d_lya_major}). All of the combinations of transitions reported in ref.~\cite{pieriCompositeSpectrum2010} are shown in lines with alternating styles and colors.}
    \label{fig:xi1d_lya_all}
\end{figure}

Figure~\ref{fig:xi1d_lya_all} shows all of the combinations of transitions reported in ref.~\cite{pieriCompositeSpectrum2010}. This is otherwise the same as figure~\ref{fig:xi1d_lya_major}.
There are interesting hints at correlations between O~\textsc{i} (1302~\AA) and C~\textsc{ii} (1335~\AA) manifested as a minor peak $v=7.4\times 10^3$\kms, and between Si~\textsc{ii} (1260~\AA) and O~\textsc{i} (1302~\AA) manifested as a minor peak $v=9.7\times 10^3$\kms.

\begin{figure}
    \centering
    \includegraphics[width=0.9\linewidth]{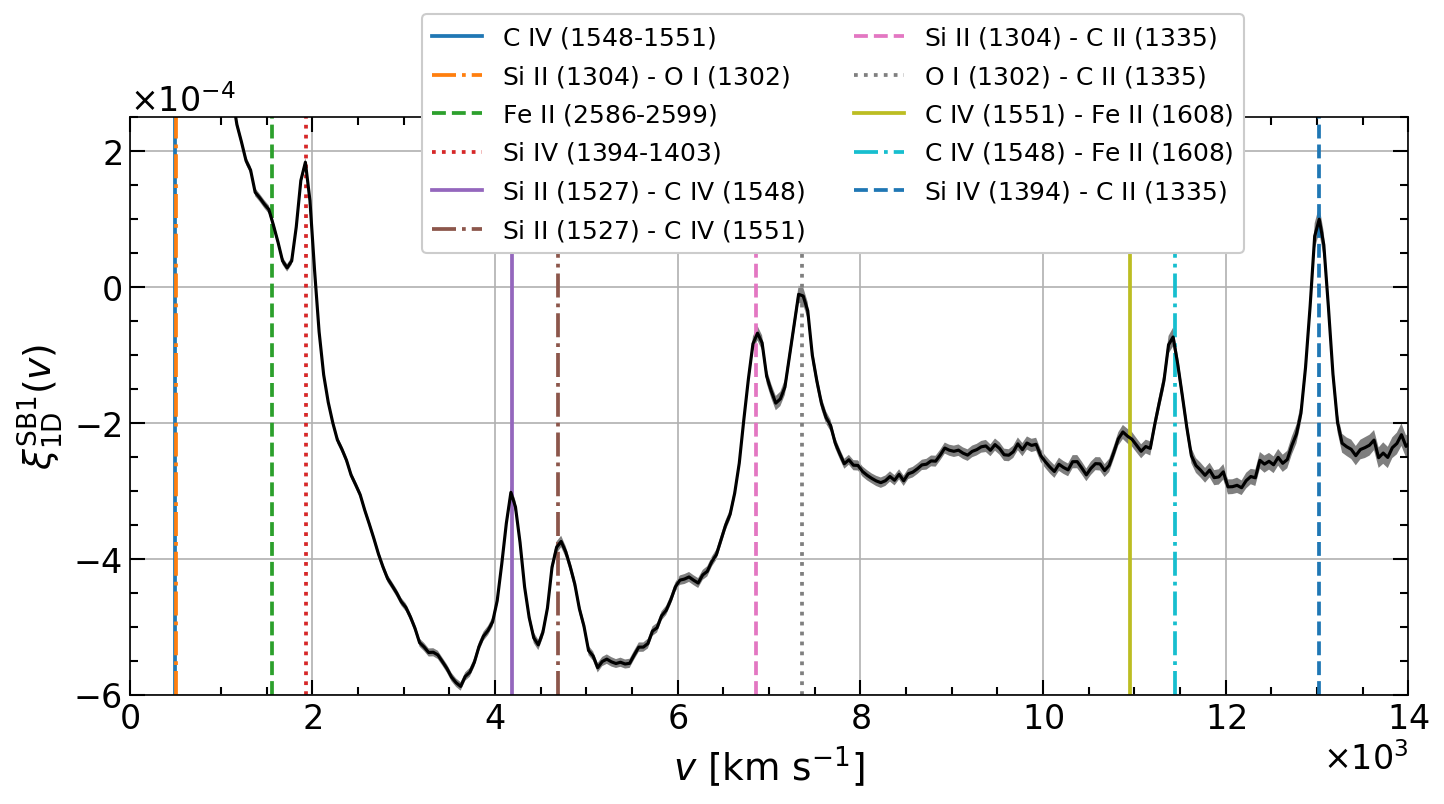}
    \caption{The weighted average of the 1D correlation function $\xi_\mathrm{1D}$ across redshift bins measured with SB1. All the combinations of transitions reported in ref.~\cite{pieriCompositeSpectrum2010} are shown in lines with alternating styles and colors.}
    \label{fig:xi1d_sb1_all}
\end{figure}

Figure~\ref{fig:xi1d_sb1_all} presents the weighted average $\xi_\mathrm{1D}$ across redshifts from SB1 and displays all the combinations of reported transitions.
The strong peak at $v=1.3\times 10^4~$\kms\ in the \lya\ $\xi_\mathrm{1D}$ is present in SB1 due to Si~\textsc{iv} (1394~\AA)$-$C~\textsc{ii} (1335~\AA). There is also strong evidence for correlations between singly and triply ionized silicon and carbon, and between neutral oxygen and singly ionized silicon and carbon.
The strong peak at $v=1.17\times 10^4~$\kms\ is likely coming from C~\textsc{iv}$-$Fe~\textsc{ii} correlations, which we confirm are also present in the SB2 correlation function.

\section{Cosmological model details\label{app:metal_model}}
We present the details of our metal oscillations model, nuisance parameters, and their priors in this section.

Let us start this section by providing details on our silicon model. As a reminder, the silicon lines we consider are Si~\textsc{iii}~($1206.51~$\AA), Si~\textsc{ii}~($1190.42~$\AA), Si~\textsc{ii}~($1193.28~$\AA), and Si~\textsc{ii}~($1260.42~$\AA). The pivot transition for all Si~\textsc{ii} lines is the $1260.42~$\AA\ line, so the amplitude parameter $a_{\mathrm{Si~II}}$ refers to its bias relative to the \lya\ field. Other lines are scaled with respect to oscillator strength and transition wavelength under the optically thin limit. Assuming these systems are completely correlated with the \lya\ field, the total flux fluctuations $\delta_F$ can be formulated as:
\begin{equation}
    \delta_F(v) = \delta(v) +  a_{\mathrm{Si~III}} \delta(v + \mu^{iii}) + a_{\mathrm{Si~II}} \left[\delta(v + \mu^{ii}_3) + r_2 \delta(v + \mu^{ii}_2) + r_1 \delta(v + \mu^{ii}_1)\right),
\end{equation}
where $\mu^m_j = c |\ln \lambda^m_j/\lambda_\mathrm{Ly\alpha}|$ for all transitions and $r_j = f_j \lambda_j / f_{1260} \lambda_{1260}$ for the $1190$ and $1193$ transitions of Si~\textsc{ii}. The oscillator strength values we use are $f_{1190} = 0.277, f_{1193} = 0.575, \mathrm{and~} f_{1260} = 1.22$ \cite{NIST_ASD}.

\begin{figure}
    \centering
    \includegraphics[width=0.5\linewidth]{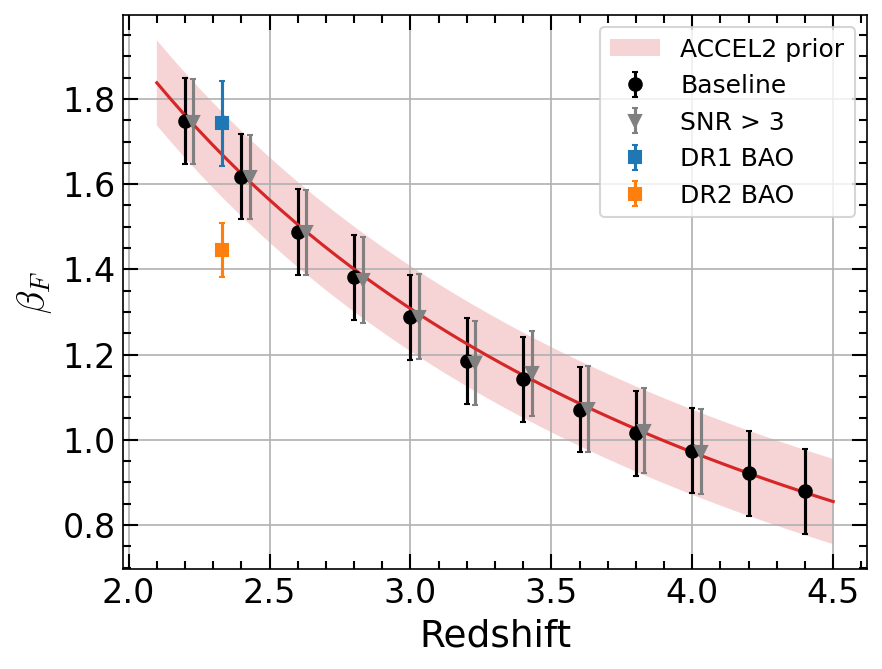}
    \caption{$\beta_F$ values are driven by the ACCEL2 prior.}
    \label{fig:accel2_beta}
\end{figure}

Going through the Fourier transform and considering all possible combinations results in the following expression for the oscillatory correction term:
\begin{equation}
    f_\mathrm{metals}(k) = C_0 + C_{\alpha m}(k) + C_{m}(k) + C_{mm}(k),
\end{equation}
where $C_0$ is the constant amplitude term summed over all metal species $m$ and their transitions:
\begin{equation}
    C_0 = \sum_{m\in \mathrm{Si~II}, \mathrm{Si~III}} a_m^2 \sum_j  \left( \frac{f_j \lambda_j}{f_m \lambda_m}\right)^2,
\end{equation}
where $f_m, \lambda_m$ refer to the pivot values (relevant only for $m=\mathrm{Si~II}$).
The $C_{\alpha m}(k)$ is the dominant oscillation term due to the offset from the \lya\ line:
\begin{equation}
    C_{\alpha m}(k) = 2 \sum_{m, j} a_m \frac{f_j \lambda_j}{f_m \lambda_m} \cos(k\mu^m_j).
\end{equation}
The third and fourth terms are the addition of our work. They account for the oscillations arising from inter-metal line combinations. The $C_m(k)$ term accounts for the oscillations coming from the different lines of a single metal species:
\begin{equation}
    C_{m}(k) = \sum_{m} a_m^2 \sum_{i, j \neq i} \frac{f_i \lambda_if_j \lambda_j}{(f_m \lambda_m)^2} \cos(k\mu^m_{ij}),
\end{equation}
where $\mu^m_{ij} = c |\ln \lambda^m_i/\lambda^m_j|$. The last term, $C_{mm}(k)$ accounts for  the oscillations coming from different metal species:
\begin{equation}
    C_{mm}(k) = \sum_{m, i} a_m \frac{f_i \lambda_i}{f_m \lambda_m} \sum_{m'\neq m, j} a_{m'} \frac{f_j \lambda_j}{f_{m'} \lambda_{m'}} \cos(k\mu^{mm'}_{ij}),
\end{equation}
where $\mu^{mm'}_{ij} = c |\ln \lambda^m_i/\lambda^{m'}_j|$. Finally, since these metal systems decorrelate at small scales, we damp these oscillations with a Gaussian function as described in the main text, such that the final correction is: $1 + f_\mathrm{metals}(k) \exp(-k^2/2k_s^2)$ and $k_s=0.009~$\skm. Freeing this $k_s$ parameter does not improve the fit.

\begin{figure}
    \centering
    \includegraphics[width=\linewidth]{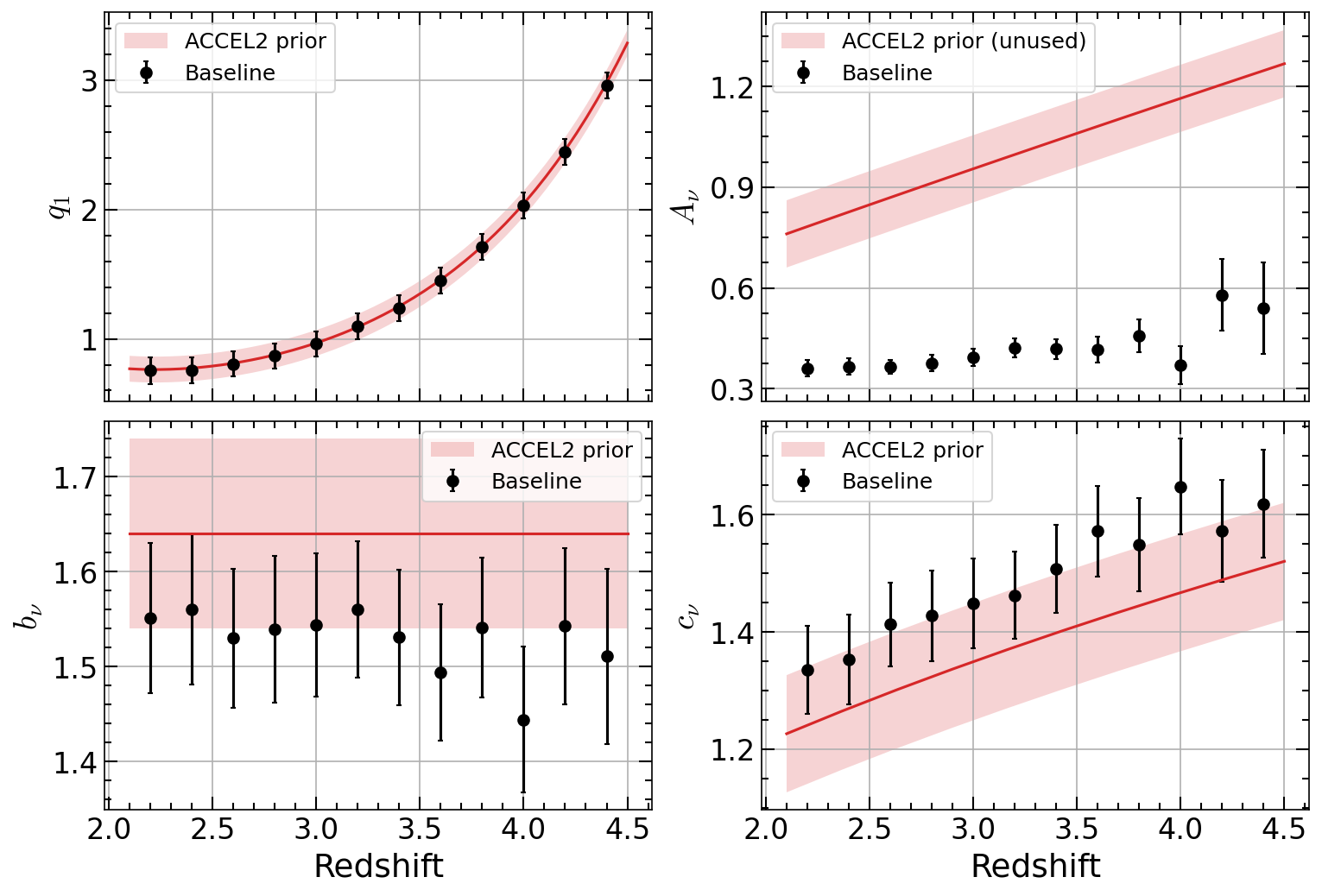}
    \caption{ACCEL2 priors and minimizer findings. $A_\nu$ prior is strongly rejected by data, so it is not used. Other parameters remain consistent with ACCEL2 priors. The best-fitting power law to $A_\nu$ is given in the text.}
    \label{fig:accel2_priors}
\end{figure}

Figure~\ref{fig:accel2_beta} shows the consistency of the $\beta_F$ values with the ACCEL2 prior. $\beta_F$ results from both baseline and high-SNR samples are driven by the prior across all redshifts, which is consistent with the DR1 BAO measurement but inconsistent with DR2. 

Figure~\ref{fig:accel2_priors} shows the nuisance parameters and ACCEL2 priors. $A_\nu$ prior from ACCEL2 is strongly rejected by data, so the prior is not applied in fitting. Other parameters remain consistent with ACCEL2 priors, although $b_\nu$ tends to prefer smaller values and $c_\nu$ larger values. Removing priors on any of these (except, of course, $A_\nu$) pushes the parameter to the boundary and results in invalid minimums. The best-fitting power law to $A_\nu$ is
\begin{equation}
    A_\nu(z) = (0.396\pm0.009)\left(\frac{1 + z}{4}\right)^{0.53\pm0.18}.
\end{equation}


\section{Author Affiliations}
\label{sec:affiliations}




\noindent \hangindent=.5cm $^{4}${Lawrence Berkeley National Laboratory, 1 Cyclotron Road, Berkeley, CA 94720, USA}

\noindent \hangindent=.5cm $^{5}${Department of Physics, Boston University, 590 Commonwealth Avenue, Boston, MA 02215 USA}

\noindent \hangindent=.5cm $^{6}${IRFU, CEA, Universit\'{e} Paris-Saclay, F-91191 Gif-sur-Yvette, France}

\noindent \hangindent=.5cm $^{7}${Dipartimento di Fisica ``Aldo Pontremoli'', Universit\`a degli Studi di Milano, Via Celoria 16, I-20133 Milano, Italy}

\noindent \hangindent=.5cm $^{8}${INAF-Osservatorio Astronomico di Brera, Via Brera 28, 20122 Milano, Italy}

\noindent \hangindent=.5cm $^{9}${Department of Physics \& Astronomy, University College London, Gower Street, London, WC1E 6BT, UK}

\noindent \hangindent=.5cm $^{10}${Institut de F\'{i}sica d’Altes Energies (IFAE), The Barcelona Institute of Science and Technology, Edifici Cn, Campus UAB, 08193, Bellaterra (Barcelona), Spain}

\noindent \hangindent=.5cm $^{11}${NASA Einstein Fellow}

\noindent \hangindent=.5cm $^{12}${Instituto de F\'{\i}sica, Universidad Nacional Aut\'{o}noma de M\'{e}xico,  Circuito de la Investigaci\'{o}n Cient\'{\i}fica, Ciudad Universitaria, Cd. de M\'{e}xico  C.~P.~04510,  M\'{e}xico}

\noindent \hangindent=.5cm $^{13}${NSF NOIRLab, 950 N. Cherry Ave., Tucson, AZ 85719, USA}

\noindent \hangindent=.5cm $^{14}${Department of Astronomy \& Astrophysics, University of Toronto, Toronto, ON M5S 3H4, Canada}

\noindent \hangindent=.5cm $^{15}${Department of Physics \& Astronomy and Pittsburgh Particle Physics, Astrophysics, and Cosmology Center (PITT PACC), University of Pittsburgh, 3941 O'Hara Street, Pittsburgh, PA 15260, USA}

\noindent \hangindent=.5cm $^{16}${University of California, Berkeley, 110 Sproul Hall \#5800 Berkeley, CA 94720, USA}

\noindent \hangindent=.5cm $^{17}${Departamento de F\'isica, Universidad de los Andes, Cra. 1 No. 18A-10, Edificio Ip, CP 111711, Bogot\'a, Colombia}

\noindent \hangindent=.5cm $^{18}${Observatorio Astron\'omico, Universidad de los Andes, Cra. 1 No. 18A-10, Edificio H, CP 111711 Bogot\'a, Colombia}

\noindent \hangindent=.5cm $^{19}${Institut d'Estudis Espacials de Catalunya (IEEC), c/ Esteve Terradas 1, Edifici RDIT, Campus PMT-UPC, 08860 Castelldefels, Spain}

\noindent \hangindent=.5cm $^{20}${Institute of Cosmology and Gravitation, University of Portsmouth, Dennis Sciama Building, Portsmouth, PO1 3FX, UK}

\noindent \hangindent=.5cm $^{21}${Institute of Space Sciences, ICE-CSIC, Campus UAB, Carrer de Can Magrans s/n, 08913 Bellaterra, Barcelona, Spain}

\noindent \hangindent=.5cm $^{22}${Fermi National Accelerator Laboratory, PO Box 500, Batavia, IL 60510, USA}

\noindent \hangindent=.5cm $^{23}${Steward Observatory, University of Arizona, 933 N. Cherry Avenue, Tucson, AZ 85721, USA}

\noindent \hangindent=.5cm $^{24}${Institut d'Astrophysique de Paris. 98 bis boulevard Arago. 75014 Paris, France}

\noindent \hangindent=.5cm $^{25}${Department of Physics, The University of Texas at Dallas, 800 W. Campbell Rd., Richardson, TX 75080, USA}

\noindent \hangindent=.5cm $^{26}${Department of Physics, Southern Methodist University, 3215 Daniel Avenue, Dallas, TX 75275, USA}

\noindent \hangindent=.5cm $^{27}${Department of Physics and Astronomy, University of California, Irvine, 92697, USA}

\noindent \hangindent=.5cm $^{28}${Sorbonne Universit\'{e}, CNRS/IN2P3, Laboratoire de Physique Nucl\'{e}aire et de Hautes Energies (LPNHE), FR-75005 Paris, France}

\noindent \hangindent=.5cm $^{29}${Departament de F\'{i}sica, Serra H\'{u}nter, Universitat Aut\`{o}noma de Barcelona, 08193 Bellaterra (Barcelona), Spain}

\noindent \hangindent=.5cm $^{30}${Instituci\'{o} Catalana de Recerca i Estudis Avan\c{c}ats, Passeig de Llu\'{\i}s Companys, 23, 08010 Barcelona, Spain}

\noindent \hangindent=.5cm $^{31}${Department of Mathematics and Theory, Peng Cheng Laboratory, Shenzhen, Guangdong 518066, China}

\noindent \hangindent=.5cm $^{32}${Departamento de F\'{\i}sica, DCI-Campus Le\'{o}n, Universidad de Guanajuato, Loma del Bosque 103, Le\'{o}n, Guanajuato C.~P.~37150, M\'{e}xico}

\noindent \hangindent=.5cm $^{33}${Instituto Avanzado de Cosmolog\'{\i}a A.~C., San Marcos 11 - Atenas 202. Magdalena Contreras. Ciudad de M\'{e}xico C.~P.~10720, M\'{e}xico}

\noindent \hangindent=.5cm $^{34}${Kavli Institute for Astronomy and Astrophysics at Peking University, PKU, 5 Yiheyuan Road, Haidian District, Beijing 100871, P.R. China}

\noindent \hangindent=.5cm $^{35}${Department of Physics and Astronomy, University of Waterloo, 200 University Ave W, Waterloo, ON N2L 3G1, Canada}

\noindent \hangindent=.5cm $^{36}${Perimeter Institute for Theoretical Physics, 31 Caroline St. North, Waterloo, ON N2L 2Y5, Canada}

\noindent \hangindent=.5cm $^{37}${Waterloo Centre for Astrophysics, University of Waterloo, 200 University Ave W, Waterloo, ON N2L 3G1, Canada}

\noindent \hangindent=.5cm $^{38}${Aix Marseille Univ, CNRS, CNES, LAM, Marseille, France}

\noindent \hangindent=.5cm $^{39}${Instituto de Astrof\'{i}sica de Andaluc\'{i}a (CSIC), Glorieta de la Astronom\'{i}a, s/n, E-18008 Granada, Spain}

\noindent \hangindent=.5cm $^{40}${Departament de F\'isica, EEBE, Universitat Polit\`ecnica de Catalunya, c/Eduard Maristany 10, 08930 Barcelona, Spain}

\noindent \hangindent=.5cm $^{41}${Universit\'{e} Clermont-Auvergne, CNRS, LPCA, 63000 Clermont-Ferrand, France}

\noindent \hangindent=.5cm $^{42}${Department of Physics and Astronomy, Sejong University, 209 Neungdong-ro, Gwangjin-gu, Seoul 05006, Republic of Korea}

\noindent \hangindent=.5cm $^{43}${CIEMAT, Avenida Complutense 40, E-28040 Madrid, Spain}

\noindent \hangindent=.5cm $^{44}${Max Planck Institute for Extraterrestrial Physics, Gie\ss enbachstra\ss e 1, 85748 Garching, Germany}

\noindent \hangindent=.5cm $^{45}${Department of Physics, University of Michigan, 450 Church Street, Ann Arbor, MI 48109, USA}

\noindent \hangindent=.5cm $^{46}${Department of Physics \& Astronomy, Ohio University, 139 University Terrace, Athens, OH 45701, USA}

\noindent \hangindent=.5cm $^{47}${Instituto de Astrof\'{\i}sica de Canarias, C/ V\'{\i}a L\'{a}ctea, s/n, E-38205 La Laguna, Tenerife, Spain}

\noindent \hangindent=.5cm $^{48}${Graduate Institute of Astrophysics and Department of Physics, National Taiwan University, No. 1, Sec. 4, Roosevelt Rd., Taipei 10617, Taiwan}

\noindent \hangindent=.5cm $^{49}${Excellence Cluster ORIGINS, Boltzmannstrasse 2, D-85748 Garching, Germany}

\noindent \hangindent=.5cm $^{50}${University Observatory, Faculty of Physics, Ludwig-Maximilians-Universit\"{a}t, Scheinerstr. 1, 81677 M\"{u}nchen, Germany}

\noindent \hangindent=.5cm $^{51}${Institute of Physics, Laboratory of Astrophysics, \'{E}cole Polytechnique F\'{e}d\'{e}rale de Lausanne (EPFL), Observatoire de Sauverny, Chemin Pegasi 51, CH-1290 Versoix, Switzerland}

\noindent \hangindent=.5cm $^{52}${National Astronomical Observatories, Chinese Academy of Sciences, A20 Datun Road, Chaoyang District, Beijing, 100101, P.~R.~China}

\bibliographystyle{JHEP}
\bibliography{references}
\end{document}